\documentclass[amsmath,amssymb,epsfig,preprint,showpacs]{revtex4}

\usepackage{graphicx}
\usepackage{dcolumn}
\usepackage{bm}
\begin{document}

\title{Interactions of Parametrically Driven Dark Solitons. II: \\
N\'eel-Bloch interactions}
\author{I.V. Barashenkov}
 \email{Igor.Barashenkov@uct.ac.za; igor@odette.mth.uct.ac.za}
\affiliation{Department of Physics, University of Bayreuth, D-95440
Bayreuth, Germany}
\altaffiliation{On sabbatical leave from
University of Cape
Town. Permanent address: Department of Applied Mathematics,
University of Cape
Town, Rondebosch 7701, South Africa}
\author{S.R. Woodford}
 \email{s.woodford@fz-juelich.de}
\affiliation{Theorie I, Institut f\"ur Festk\"orperforschung,
Forschungszentrum J\"ulich,  D-52428 J\"ulich, Germany}

\date{\today}

\begin{abstract}
 The interaction between a Bloch and a N\'eel wall
 in the parametrically driven nonlinear Schr\"odinger equation is
 studied by following the dissociation of their unstable bound
state.
Mathematically, the analysis focusses on the splitting of a four-fold  zero eigenvalue
associated with a pair of infinitely separated Bloch and
N\'eel walls.
It is shown that a Bloch and a N\'eel wall interact as two classical
particles, one with positive and the other one with negative mass.

\end{abstract}


\maketitle

\section{Introduction}
\label{Intro}

In the preceding publication \cite{BWZ1} we
started our analysis of the interactions between
dark solitons of the parametrically driven nonlinear
Schr\"odinger (NLS) equation:
\begin{equation}
\label{NLS}
i\partial_T \Psi + {\textstyle \frac 12} \partial_{X}^2 \Psi
+ \Psi - |\Psi|^2\Psi = h\Psi^* - i\gamma\Psi.
\end{equation}
This equation arises in a wide variety of physical contexts;
see \cite{BWZ1} for references.
In Eq.(\ref{NLS}),
$h$ is the strength of the parametric driving and $\gamma$ is
the damping coefficient. In the nondissipative
limit, when $\gamma=0$, the equation has two coexisting stable soliton
solutions, the Bloch and the N\'eel wall.
In \cite{BWZ1},
we considered forces existing between two Bloch walls
and between two N\'eel walls.
The present work completes the analysis by classifying
the N\'eel-Bloch interactions. The understanding of this nonsymmetric
situation requires a mathematical formalism different
from the one used in \cite{BWZ1}; this justifies the need for a
separate treatment. Since the Bloch
wall does not exist for $\gamma \neq 0$, we consider here the nondissipative
case only. This is
another distinction from  Ref.\cite{BWZ1}.

Our analysis of the
interaction between Bloch and N\'eel walls
will be based on the study of linearised perturbations
about their (unstable) bound state.
Mathematically, this analysis reduces to the
construction of eigenfunctions of a Schr\"odinger-like
operator, acting in the space of vector-functions,
with the potential consisting of two well-separated
nonidentical potential wells. When the two wells are
infinitely far apart, there are four zero eigenvalues
in its spectrum, with two associated eigenfunctions.
As the wells (produced by the Bloch and N\'eel walls)
are moved closer together, the degeneracy is partially lifted,
with only two eigenvalues remaining at the origin.
The question here is whether the two nonzero eigenvalues
move onto the imaginary axis --- which would be the case of
stability, or onto the real axis --- in which case the
bound state is unstable.
We use matched asymptotic expansions to show that the
second is the case and construct
eigenfunctions associated with the two real eigenvalues.

The eigenfunction associated with the positive, unstable,
eigenvalue contains the entire information on the
character of interaction of two walls. We demonstrate that
the outcome of the N\'eel-Bloch interaction depends on
their left-right arrangement and the chirality of the Bloch wall.
 A N\'eel wall and a {\it right}-handed Bloch wall placed on its right
 will be
moving in the same direction. If, however, we place a
 {\it left}-handed Bloch wall on the right of the N\'eel, the two walls will
move in opposite
directions
--- towards or away from each other, depending on the
initial perturbation.
The
right-handed Bloch
wall (on the left) and  the N\'eel wall
(on the right) will  move in opposite directions,
while a pair involving the left-handed Bloch on the left
of the N\'eel, will move colinearly.

After the eigenfunctions associated with two opposite
real eigenvalues have been constructed, the evolution of an
arbitrary initial condition close to a pair of well separated
Bloch and N\'eel walls is not difficult to predict. Treating this
initial condition as a perturbed Bloch-N\'eel
bound state, its evolution  will be determined  by
 the projection of the  perturbation on the bubble's
  unstable eigendirection. We illustrate this
general approach
by considering an example of initial condition in the form of a
product of the Bloch and N\'eel wall.

Usually one tries to understand the interaction of solitons as
interaction of point-like particles; the particle description
is physically appealing and mathematically lucid.
We will show that a Bloch and a N\'eel wall {\it can\/} be treated
as  two classical particles. However the interaction between these
two particles is anomalous in the sense that the Bloch wall being attracted to
the N\'eel wall does not
necessarily imply the reciprocal attraction of the
N\'eel  to the Bloch. This anomaly can be
understood by considering the N\'eel wall as a particle with
negative mass. The ``wrong" mass sign arises very naturally if one
recalls what the N\'eel wall really is: a localised depression, a patch of
low density moving over a high-density background. The only reason
why this property was not fully appreciated before is because earlier studies focussed on
symmetric, N\'eel-N\'eel, interactions --- which are, of course, non-anomalous.

The outline of this paper is as follows.
In section \ref{Bubbles!} we introduce {\it travelling\/}
Bloch and N\'eel walls, and describe
 the Bloch-N\'eel bound state.
 Section \ref{Splitting}  contains  the main mathematical result
 of this paper, the asymptotic analysis of the splitting of
 the degenerate zero eigenvalue.  In the next section
 (section \ref{Interpretation}) we interpret the resulting eigenfunctions
in terms of motions of the constituent walls.
 In section \ref{Interaction} we describe how the eigenfunctions can be used to
 classify the interaction of a pair of Bloch and N\'eel
 walls  and apply this approach
 to a characteristic example.
 Finally, the main
results are summarised  in section \ref{Conclusions} where we also
interpret the interaction of the walls as interaction of opposite
mass-sign particles.

\section{Moving Bloch and N\'eel
walls and the Bloch-N\'eel Complex}
\label{Bubbles!}

In this paper, we restrict ourselves to the
undamped situation, $\gamma=0$.
As in \cite{BWZ1},
we let
\begin{equation}
\Psi(X,T) = iA \psi(x,t), \quad x = AX, \quad t=A^2 T,
\label{MsEllen}
\end{equation}
with
\begin{equation}
A = \sqrt{1 + h}.
\end{equation}
Equation (\ref{NLS}) becomes
\begin{equation}
i \psi_t + {\textstyle \frac {1}{2}} \psi_{xx} -
 |\psi|^2 \psi + {\textstyle \frac {1}{A^2}} \psi  +  {\textstyle \frac {h}{A^2}} \psi^*
= 0.
\label{NLS2}
\end{equation}
This
is the form of the parametrically driven
 NLS that will be used in this paper.
The stable background solutions of Eq.(\ref{NLS2}) are
$\psi_{\mbox{flat}}= \pm 1$.
 Without loss of generality we assume that $h>0$.

The two topological solitons of (\ref{NLS2}) were
introduced in Ref.\cite{BWZ1}. One is
 the  N\'eel wall \cite{XY,Raj,Niez,Niez2,Elphick_Meron}:
\begin{equation}
\label{Neel}
\psi_N  (x) =  - \tanh(x).
\end{equation}
Note that we are introducing the N\'eel wall differently from \cite{BWZ1},
with an extra negative sign in front of the $\mbox{tanh}$. This is
done for  later convenience.
(In Ref.\cite{BWZ1}, we
would refer to the solution (\ref{Neel}) as an {\it antiwall}.)

The second topological soliton
is   the
Bloch wall \cite{Sarker,Montonen,Niez,Niez2}:
\begin{equation}
\label{Bloch}
\psi_B (x) = \tanh(Bx) \pm i \sqrt{1-B^2} \,   \mbox{sech}(Bx),
\end{equation}
where
\[
  B = 2\frac{\sqrt{A^2-1} }{A} = 2 \sqrt{ \frac{h}{1+h}}.
\]
Equation (\ref{Bloch}) with the positive sign
in front of the imaginary part describes the
right-handed
Bloch wall  while in the case of the
negative sign, the wall is said to be left-handed (see \cite{OurPaper}
for details).

The N\'eel wall exists for all $h>0$ whereas the
Bloch wall exists only for $A^2 < \frac43$, i.e. for $0<h< \frac13$.
Since we are interested in the Bloch-N\'eel interaction,
the latter will be our region of consideration.
Both the Bloch and N\'eel walls are stable in their entire regions
of existence,
$0<h< \frac13$ and $h>0$, respectively \cite{OurPaper}.

The Bloch and N\'eel
walls can be continued to nonzero velocity
for all $h$\ values
for which they exist \cite{OurPaper}. The moving walls
of the form $\psi=\psi(x-vt)$ are found
as solutions of the ordinary differential
equation
\begin{equation}
-i v \psi_x + {\textstyle \frac {1}{2}} \psi_{xx} -
 |\psi|^2 \psi + {\textstyle \frac {1}{A^2}} \psi  +  {\textstyle \frac {h}{A^2}} \psi^*
= 0.
\label{stationary}
\end{equation}
The solution obtained by the continuation from the
stationary N\'eel wall $\psi_N(x)$, will be referred to as the
``moving N\'eel wall" and denoted $\psi_N(v;x)$.
In a similar way, the solution obtained by continuing $\psi_B(x)$
in $v$, will be called the ``moving Bloch wall";
to be denoted  $\psi_B(v;x)$.
 The travelling walls are stable for all velocities $v$ \cite{OurPaper}.

\begin{figure}
\includegraphics[ height = 2.1in, width = 0.5\linewidth]{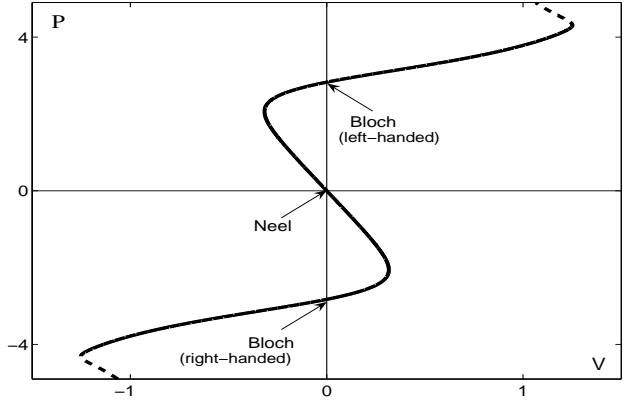}
\caption{\sf The bifurcation diagram for the stationary and
travelling Bloch and N\'eel walls with $h<\frac13$
(adapted from Ref.\cite{OurPaper}). The momentum
(\ref{momento}) is used as the bifurcation  measure.
 Solid curve
indicates stable and dashed curve unstable solutions.
Note that the mass of the N\'eel wall (defined as $dP/dv$) is
opposite to the mass of the Bloch wall.
In this plot, $h=0.05$.
}
\label{momentum_vs_V}
\end{figure}

An important characteristic of solutions of Eq.(\ref{NLS2})
is their field momentum:
\begin{equation}
P= \frac{i}{2}  \int (\psi \psi_x^* - \psi^* \psi_x) dx.
\label{momento}
\end{equation}
The momentum is conserved: $dP/dt=0$.
When we take the
complex conjugate of $\psi(x,t)$, the associated momentum
changes its sign: $P[\psi^*]=-P[\psi]$. Consequently, the momenta of two stationary
Bloch walls  with opposite chiralities are opposite. The right-handed
 Bloch wall has
negative momentum, while the left-handed wall's momentum is positive.
The momentum of the stationary N\'eel wall is, naturally, equal to zero.
Figure \ref{momentum_vs_V}
 shows the momenta of stationary and travelling
Bloch and N\'eel walls.
These will be denoted  $P_B$ and $P_N$, respectively:
\begin{subequations}
\begin{eqnarray}
P_B= P_B(v) \equiv P[\psi_B(v;x)],
\label{yB} \\
P_N= P_N(v) \equiv P[\psi_N(v;x)].
\label{yN}
\end{eqnarray}
\end{subequations}

According to Fig.\ref{momentum_vs_V}, when the right-handed
Bloch wall is continued  into the region $v>0$, it transforms into
the moving N\'eel wall. In a similar way, when we path-follow
the left-handed Bloch wall to large negative velocities,
the corresponding branch turns into the branch of travelling
N\'eel walls. Thus, the classification of moving solutions into
Bloch and N\'eel walls is only sensible for
sufficiently small $v$; for higher velocities,
there is no qualitative difference between the two types of walls.

One more observation with regard to Fig.\ref{momentum_vs_V} concerns
particle properties of the two walls. While the stationary Bloch wall has
positive mass, $m_B \equiv (dP_B/dv)_{v=0}>0$, the mass of the N\'eel
wall is negative: $m_N \equiv (dP_N/dv)_{v=0}<0$. This property will
be crucial for the particle interpretation of the N\'eel-Bloch interaction
(section \ref{Conclusions}).

In addition to
the Bloch and N\'eel  domain walls,
Eq.(\ref{NLS2}) possesses nontopological solitons.
One such solution,
arising for $h = \frac{1}{15}$,
is well known \cite{Raj2, SubTrull, Sarker,
Hawrylak}:
\begin{equation}
\label{particular}
\psi = 1 -  \frac 32 \mbox{sech}^2 \left(\frac x2\right) \pm   \frac{3i}{2}
\tanh \left( \frac x2 \right) \mbox{sech} \left( \frac x2 \right).
\end{equation}
Recently, it has become clear that this solution is a member of a one-parameter
family of solutions which exist for all $h < \frac 13$\  \cite{OurPaper2}.
For each $h$, this family has the following analytical expression:
\begin{subequations}
\label{Family}
\begin{equation}
\psi_b = \frac{p + i \sigma q}{\cal D},
\label{FamilyA}
\end{equation}
where
\begin{eqnarray}
p = 1 - e^{\phi_1 +  2\beta} - e^{2\phi_2 + 2\beta} + e^{\phi_1 + 2\phi_2},
\label{FamilyB} \\
q = 2(1+B)e^{\phi_2}(1 - e^{\phi_1}),
\label{FamilyC}
\end{eqnarray}
and
\begin{equation}
{\cal D} = 1 + e^{\phi_1 + 2\beta} + e^{2\phi_2 + 2\beta} + e^{\phi_1 + 2\phi_2}.
\label{FamilyD}
\end{equation}
\end{subequations}
In Eq.(\ref{Family}), $\sigma$ is a sign factor:  $\sigma = \pm
1$; the exponents $\phi_1$, $\phi_2$ are
\begin{equation}
 \phi_1 = 2(x+s), \quad \phi_2 = B(x-s),
\label{Parameters}
  \nonumber
\end{equation}
and  $\beta$ is defined by
\[
 B \equiv {\rm tanh\/} \, \beta.
\]
The solution (\ref{Family}) describes a
bound state, or a complex,  of a Bloch and a N\'eel wall, with the parameter $s$\ characterising
the distance between the centres of the two walls \cite{OurPaper2}.
The choice $s>0$\ corresponds to the Bloch wall on the right of the
N\'eel wall; for $s<0$, the Bloch wall is on the left.
The solution (\ref{particular}) corresponds to $s=0$ (and $h=1/15$).
  The sign factor $\sigma$\ determines the chirality of the Bloch wall bound in the complex:
$\sigma = 1$\ implies that
the Bloch wall is left-handed, while  $\sigma = -1$\
identifies a complex with a right-handed Bloch wall.

 Eq.(\ref{Family}) can also be interpreted as a bubble of one
phase ($\psi = -1$) embedded in a background of a different phase ($\psi = 1$);
we will frequently be referring to this solution as a ``bubble''
\cite{bubbles}.

The analysis of the bound
state (\ref{Family}) will
provide the understanding of the Bloch-N\'eel interaction.

\section{Splitting of the degenerate zero modes}
\label{Splitting}

\subsection{Stability problem for the complex}

To study the stability of the bubble  (\ref{Family}),
as well as to explore the phase space in the neighbourhood of this solution,
we linearise
Eq.(\ref{NLS2}) in the small perturbation $\delta\psi (x,t)$.
Assuming the time dependence of the form
$\delta \psi(x,t)= \left[ u(x) +i w(x) \right] e^{\lambda  t}$,
  results in the eigenvalue problem
\begin{equation}
\label{EVprob}
\mathcal{H} \vec{\varphi} = \lambda J \vec{\varphi},
\end{equation}
where $\vec{\varphi}$ is a 2-vector made of the real
and imaginary parts of the perturbation:
\[
\vec{\varphi} = \left(\begin{array}{c} u \\ w \end{array} \right),
\]
$J$ is a skew-symmetric matrix
\begin{equation}
 J = \left(\begin{array}{lr} 0 & -1 \\ 1 & 0 \end{array} \right),
 \label{vectorthings}
\end{equation}
and $\mathcal{H}$\ is a hermitian operator:
\begin{eqnarray}
\mathcal{H} =
-\frac I2 \partial_x^2 +  \left(\begin{array}{lr}
 3 {\cal R}^2 +  {\cal I}^2 - 1 & 2 {\cal R} {\cal I} \\
2  {\cal R} {\cal I} &  {\cal R}^2 + 3 {\cal I}^2 -
\frac{2 - A^2}{A^2} \end{array} \right).
\nonumber \\
\label{LinearisedOperator}
\end{eqnarray}
In the last equation,
 $I$ stands for the $ 2 \times 2$ identity matrix, and ${\cal R}$\ and ${\cal I}$\ are
the real and imaginary parts of the solution (\ref{Family}): $\psi_b = {\cal R}(x) + i{\cal I}(x)$.
In what follows, we restrict our attention to the bound state with
the N\'eel wall on the left of the Bloch wall [$s > 0$\ in Eq.(\ref{Family})].
The results for $s < 0$\ will be recovered by exploiting the
symmetry $\psi(-x;-s) = \psi^*(x;s)$\ of the solution
(\ref{Family}). Since the dynamics described by equation (\ref{NLS2})
are invariant under the reflection $x\to -x$, the
evolution of a bubble with $s = s_0 <0$\ will follow the same pattern as the evolution of
the conjugate bubble with positive $s = -s_0$.

For all values of $s$, the continuous spectrum of the operator
$\mathcal{H}$\ lies on the imaginary axis, with $|\mbox{Im} \, \lambda | >  B$, and does not
give rise to instabilities.
For $s= \infty$, the separation of the Bloch and N\'eel wall in the
bubble is infinite and so we have essentially two
independent stationary (but potentially mobile) walls, each having
two zero eigenvalues in its linearised spectrum. One of these stems from
the
translation invariance while the other one is associated with velocity boosts of the
corresponding wall.
For finite $s$, only two of the four  zero eigenvalues remain in the spectrum:
one pertaining to the translation invariance of the
complex as a whole and the other one associated with variations of the
interwall separation. In this section we
compute, perturbatively, the arising {\it nonzero\/}
eigenvalues (and hence classify the stability of the bubble).
We also construct the eigenfunctions associated with the real
eigenvalues  --- these will provide insight into the evolution of
the unstable bound state and nearby initial conditions.

The lifting of the degeneracy of a repeated eigenvalue of the
scalar Schr\"odinger operator with the potential comprising
 two identical potential wells with large separation, is discussed in the
classical textbook \cite{Landau}. This analysis is not helpful in our case,
unfortunately, for three reasons: (i) our ${\cal H}$ operates on vector, not scalar,
functions; (ii) the potential wells formed by the Bloch and N\'eel walls
are  not identical; (iii) the analysis in \cite{Landau}
 {\it postulates\/} a particular form
of the wavefunction
on  symmetry grounds,
rather than deriving it within some perturbation formalism
--- as a result, the generalisation to the vector nonsymmetric case is not
straightforward.

Our treatment will be based on expanding the
eigenfunction in the asymptotic
series near  the cores of the two walls and matching the  resulting expansions
in the overlap region $x \sim 0$. This approach builds on the asymptotic procedure used for the study
of the stability of the travelling
dark soliton  \cite{Dark_stability}.

\subsection{Left expansion.}

We will do all our calculations for the case $s>0$ (i.e. for the
Bloch wall on the right of the N\'eel wall). Furthermore, we
will restrict ourselves to  the
right-handed Bloch wall only
($\sigma=-1$). The other three possible combinations of $\sigma$
and $\mbox{Sign}(s)$ will be commented upon at the end of
section \ref{Splitting}.

Let, first, $x \in (-\infty, x_0)$, where $x_0>0$
is a fixed  value independent of $s$ (so that $x_0<s$). In this region the real
and imaginary parts of the bubble solution (\ref{Family}) can be written
as
\begin{eqnarray}
{\cal R} = -{\rm tanh} \, \xi +
\frac{ \sinh [2(\xi-\beta)]}{\cosh^2 \xi}  e^{2B(\xi-2s-\beta)} & +  & ...,
\nonumber
\\
{\cal I} =2 ({\rm tanh} \xi -B)  \, e^{B(\xi-2s-\beta)} & +  & ...,
\label{Z1}
\end{eqnarray}
where
\begin{equation}
\xi \equiv x+s +\beta,
\label{1star}
\end{equation}
 and we have dropped terms of order
$e^{3B(\xi-2s-\beta)}$ and smaller.
Equations (\ref{Z1}) can be seen as expansions of
two functions of $\xi$, defined for $-\infty< \xi< \infty$,
 in powers
of $\varepsilon=e^{-2Bs}$. Accordingly, the operator ${\cal H}$ in
Eq.(\ref{LinearisedOperator}) expands as
\begin{equation}
{\cal H}= {\cal H}_N  + \varepsilon {\cal H}_N^{(1)} + \varepsilon^2 {\cal H}_N^{(2)}+
...
\label{Z2}
\end{equation}
Here, ${\cal H}_N$ is the unperturbed linearised operator
of the N\'eel wall centred at $\xi=0$, i.e. Eq.(\ref{LinearisedOperator})
with ${\cal R} =-{\rm tanh} \, \xi$ and ${\cal I}=0$.
Guided by the results of (our own) numerical analysis, we {\it assume\/}
that $\lambda$ is of order $\varepsilon$:
\[
\lambda=\lambda_0 \varepsilon.
\] (As we will see,
this asumption leads to a self-consistent perturbation scheme.)
This implies that  the eigenfunction ${\vec \varphi}$ in the eigenvalue
problem (\ref{EVprob}) can also be expanded in powers of $\varepsilon$:
\begin{equation}
{\vec \varphi}(\xi)= a {\vec \psi}_N'(\xi)  + \varepsilon {\vec \varphi}_1(\xi) +
\varepsilon^2 {\vec \varphi}_2(\xi)+ ...
\label{Z3}
\end{equation}
Here ${\vec \psi}_N'   \equiv \partial_\xi {\vec \psi}_N$, and ${\vec \psi}_N=({\cal R}_N,
{\cal I}_N)=(-{\rm tanh\/} \, \xi, 0) $.
The coefficient $a$ is arbitrary at this stage [$a={\cal O}(1)$].

Substituting (\ref{Z2})-(\ref{Z3}) into (\ref{EVprob}) and equating
coefficients of like powers of $\varepsilon$, yields, at order
$\varepsilon^1$:
\begin{equation}
{\cal H}_N  {\vec \varphi}_1 +a {\cal H}_N^{(1)}  {\vec \psi}_N'= a \lambda_0 J
{\vec \psi}_N'.
\label{Z8}
\end{equation}
To solve (\ref{Z8}), we note that  ${\dot {\vec \psi}}_N$ is a generalised eigenvector
associated with the zero eigenvalue:
\begin{equation}
{\cal H}_N {\dot {\vec \psi}}_N =-J {\vec \psi}_N'.
\label{Z5}
\end{equation}
 Here $\psi_N$ is considered as a solution of the form $\psi
 =\psi(x-vt)$ of equation (\ref{stationary}), while the overdot indicates
differentiation with respect to velocity $v$ (and not time):
${\dot {\vec \psi}}_N
\equiv \left( \partial_v {\vec \psi}_N  \right)_{v=0}$.
(This will be our convention until the end of section \ref{Splitting}.)
Also, we know that ${\vec \psi}_b'  \equiv
\partial_x {\vec \psi}_b$ and
$\partial_s {\vec \psi}_b$
are  zero modes of the perturbed
operator (\ref{LinearisedOperator})
where ${\vec \psi}_b(x)=({\cal R},{\cal I})$ is
the bubble solution (\ref{Family}),
and $-\infty < x < \infty$.
Using (\ref{Z1}), we write
 \begin{eqnarray}
 {\vec \psi}_b' & = & {\vec \psi}_N' +   \varepsilon \left( {\vec \chi}_1
  +{\vec y}_1 \right)
 +\varepsilon^2 \left({\vec \chi}_2 + {\vec y}_2 \right)+..., \nonumber
\\
 \partial_s {\vec \psi}_b & =  & {\vec \psi}_N' +
\varepsilon \left( {\vec \chi}_1 - {\vec y}_1 \right) +\varepsilon^2
\left( {\vec \chi}_2 - {\vec y}_2 \right)
+...,
 \label{Z6}
\end{eqnarray}
where
\begin{eqnarray}
{\vec \chi}_1(\xi) = 2 e^{B(\xi-\beta)} {\rm sech\/}^2 \xi \left( \begin{array}{c}
 0 \\ 1
 \end{array}
 \right),
\label{Z_chi} \\
{\vec y}_1(\xi) =2 B e^{B (\xi-\beta)} ({\rm tanh\/}  \, \xi-B)
\left( \begin{array}{c}
 0 \\ 1
 \end{array}
 \right),
\label{Z10} \\
{\vec \chi}_2(\xi)=
e^{2B(\xi-\beta)} \frac{d}{d \xi} \frac{ \sinh[2(\xi-\beta)]}{\cosh^2 \xi}
   \left( \begin{array}{c}
   1 \\0  \end{array} \right), \\
   {\vec y}_2(\xi)=2B e^{2B(\xi-\beta)} \frac{ \sinh[2(\xi-\beta)]}{\cosh^2 \xi}
   \left( \begin{array}{c} 1 \\ 0
 \end{array}
 \right). \label{Zy2}
\end{eqnarray}

Substituting expansions (\ref{Z2}) and (\ref{Z6}) into ${\cal H} {\vec
\psi}_b' ={\cal H} \, \partial_s {\vec \psi}_b=0$, the order $\varepsilon^1$ gives a
useful identity:
\begin{equation}
{\cal H}_N  {\vec \chi}_1 +{\cal H}_N^{(1)} {\vec \psi}_N'=0;
\label{Z7}
\end{equation}
we  also note that
\begin{equation}
{\cal H}_N  {\vec y}_1 =0.
\label{Z70}
\end{equation}
Using (\ref{Z5}) and (\ref{Z7}), we can solve Eq.(\ref{Z8}) in
the class of functions bounded as $|\xi| \to \infty$:
\[
{\vec \varphi}_1(\xi) =-a \lambda_0 {\dot {\vec \psi}}_N + a {\vec \chi}_1.
\]

The emerging vector-function
${\vec \varphi}= a {\vec \psi}_N'  + \varepsilon {\vec \varphi}_1 +...$ decays,
exponentially,   both
as $\xi \to -\infty$ and $\xi \to +\infty$
 and hence it is intuitively clear that it {\it cannot\/}
describe the behaviour of the eigenfunction of the bound state in the region
under consideration  ($x<0$).
Indeed, the eigenfunction should include a term growing as $\xi \to +\infty$
and representing the tail of the Bloch wall situated in the region $x>0$.
Therefore we need to add to ${\vec \varphi}$  a
solution of the equation ${\cal H}_N {\vec y}=0$ which decays
as $\xi \to -\infty$ but grows as $\xi \to \infty$. This solution
is already available [see (\ref{Z70})]; it is
given just by Eq.(\ref{Z10}).
Note that the  other two linearly independent solutions of equation ${\cal H}_N
{\vec y}=0$ (other than ${\vec \psi}_N'$ and ${\vec y}_1$),
are growing as $\xi \to -\infty$; these are clearly 
not acceptable for our purposes.

Finally, the full order-$\varepsilon^1$ perturbation is
$-a \lambda {\dot {\vec \psi}}_N + a \varepsilon {\vec \chi}_1 + C_1 {\vec y}_1$, and
the eigenfunction ${\vec \varphi}$ becomes, in the region $x < x_0$ (``left region"):
\begin{equation}
{\vec \varphi}(\xi)
 = a {\vec \psi}'_N -a \lambda {\dot {\vec \psi}}_N + a \varepsilon {\vec \chi}_1 + C_1 {\vec y}_1
 + {\cal O}(\varepsilon^2),
\label{Z11}
\end{equation}
with ${\vec y}_1$ as in (\ref{Z10}) and  the constant $C_1$
 to be found from the matching condition
at a later stage.
(Here we are implicitly assuming that this constant will be
of order $\varepsilon$. If, however, it is of order $\varepsilon^2$ or higher, the
term $C {\vec y}_1$ will only appear at higher orders of the expansion.)

\subsection{Right expansion.}
Let, now $x \in (-x_0, \infty)$.
(We remind that $x_0>0$ is a fixed value independent of $s$;
$x_0< s$.)
 Here, the real and imaginary parts of the bubble
are given by
\begin{eqnarray}
{\cal R}  = {\rm tanh} \, (B \eta -\beta) -
\frac{ \sinh (2B \eta) }{\cosh^2 (B \eta-\beta)}  e^{-2(\eta+2s)}  +  ...,
\nonumber \\
{\cal I}  =  \frac{{\rm sech\/} \beta}{\cosh  (B \eta-\beta)}  -
 \frac { 2 \cosh (B \eta)}{\cosh^2 (B \eta -\beta)}
 e^{-2(\eta+2s)}  +  ...,
\label{Z120}
\end{eqnarray}
where
\begin{equation}
\eta \equiv x-s
\label{2stars}
\end{equation}
 and we have dropped
terms of order $e^{-4(\eta+2s)}$. Eqs.(\ref{Z120}) can be seen as expansions
of the functions ${\cal R} (\eta)$ and ${\cal I} (\eta)$
(with $-\infty< \eta < \infty$) in powers of $\mu=e^{-4s}$.
Accordingly, the operator ${\cal H}$ expands as
\begin{equation}
{\cal H} = {\cal H}_B  + \mu {\cal H}_B^{(1)} + \mu^2 {\cal H}_B^{(2)}+
... .
\label{Z13}
\end{equation}
Here, ${\cal H}_B$ is the unperturbed linearised operator of the Bloch
wall centred at $\eta= \beta/B$, Eq.(\ref{LinearisedOperator}) with
${\cal R}_B ={\rm tanh} \, (B \eta-\beta)$ and
${\cal I}_B ={\rm sech} \, \beta \, {\rm sech} \, (B \eta
-\beta)$.
Note that the expansion parameter $\mu$ is, in general, incommensurate with $\varepsilon$
and hence it is not {\it a priori\/} obvious what the expansion
of the eigenfunction ${\vec \varphi}(\eta)$ will be.
Letting ${\vec \varphi}(\eta) = {\vec \psi}_B' + {\hat {\vec \varphi}}$,
where ${\hat {\vec \varphi}}$ has {\it some\/} order of smallness,
and substituting into  (\ref{EVprob}), we get
\begin{equation}
{\cal H}_B {\hat {\vec \varphi}}  = \lambda_0 \varepsilon J {\vec \psi}_B',
\label{Z14}
\end{equation}
where we have dropped terms of order $\mu$ because $\mu$ is smaller
than $\varepsilon$ (and even $\varepsilon^2$). From (\ref{Z14}) it is clear that
the leading-order correction to the translation mode of the free-standing Bloch wall is of order
$\varepsilon$ (and not $\mu$ or $\mu^{1/2}$ as one might have been
tempted to think).

Recalling that ${\dot {\vec \psi}}_B(\eta)
\equiv \left( \partial_v {\vec \psi}_B \right)_{v=0}$
is a generalised eigenvector associated with the
zero eigenvalue of the free-standing Bloch wall:
\[
{\cal H}_B {\dot {\vec \psi}}_B =-J {\vec \psi}_B',
\]
the localised solution of Eq.(\ref{Z14}) is given by
 ${\hat {\vec \varphi}}  =-\lambda  {\dot {\vec \psi}}_B $, and so
(the localised part of) the
eigenfunction ${\vec \varphi}$ can be written as
\begin{equation}
{\vec \varphi}(\eta) = b {\vec \psi}_B'
- \lambda b {\dot {\vec \psi}}_B + {\cal O}(\varepsilon^2, \mu),
\label{Z16}
\end{equation}
where $b$ is an arbitrary constant of order 1.

The eigenfunction (\ref{Z16}) decays, exponentially, as $\eta \to -\infty$ and therefore, cannot
represent the $(x>0)$-behaviour of the eigenfunction of the bound state.
(For there is no connection of such an eigenfunction to the region
$x<0$ where the N\'eel wall is located.)  In order to obtain
the correct behaviour, we need to add a solution of equation ${\cal H}_B {\vec z}=0$
which decays as $\eta \to +\infty$ but grows as $\eta \to -\infty$. The equation
has only two solutions that decay as $\eta \to \infty$; one is the
translation mode
\begin{equation}
{\vec \psi}_B'=({\cal R}_B',{\cal I}_B') 
=\left( B \, {\rm sech\/}^2(B \eta-\beta),
 -B \, {\rm sech\/} \beta \, \frac{{\rm tanh\/} (B \eta-\beta)}{\cosh(B \eta-\beta)} \right),
\label{Z17}
\end{equation}
and the other one can be found by expanding the exact
null eigenfunctions of the
bubble, ${\vec \psi}_b' \equiv \partial_x {\vec \psi}_b$
and $\partial_s {\vec \psi}_b$,  in powers of $\mu$
(in the region $x>-x_0$). We have
\begin{eqnarray}
{\vec \psi}_b'= {\vec \psi}_B' + \mu \left(  {\vec z_1} -   {\vec \theta_1} \right) & + & {\cal O}(\mu^2),
\nonumber
\\
\partial_s {\vec \psi}_b
= -{\vec \psi}_B' + \mu \left(  {\vec z_1} +   {\vec \theta_1} \right) & + & {\cal O}(\mu^2),
\label{Z18}
\end{eqnarray}
where
\begin{equation}
{\vec z_1}(\eta)= \frac{2 e^{-2 \eta}}{\cosh^2(B \eta-\beta)}
\left(
\begin{array}{c}
\sinh (2 B \eta) \\
2 \cosh (B \eta)
\end{array}
\right)
\label{Z19}
\end{equation}
and
\begin{equation}
{\vec \theta_1}(\eta)= \frac{e^{-2 \eta}}{2} \frac{d}{d \eta}
\left( e^{2 \eta} {\vec z_1} \, \right).
\label{Z20}
\end{equation}
Substituting (\ref{Z13}) and (\ref{Z18}) into
${\cal H} {\vec \psi}_b'={\cal H} \partial_s {\vec \psi}_b=0$,
we obtain ${\cal H}_B {\vec z}_1=0$ which means that Eq.(\ref{Z19}) gives exactly the
solution we are looking for.

Finally, the correct behaviour of the eigenfunction ${\vec \varphi}$
in the region $x>-x_0$ (the ``right" region) is given by
\begin{equation}
{\vec \varphi}(\eta)=
b {\vec \psi}_B'-\lambda b {\dot {\vec \psi}}_B + D {\vec z}_1 +\mathcal{O}
(\varepsilon^2),
\label{Z21}
\end{equation}
where ${\vec z}_1$ is given by Eq.(\ref{Z19}) and the constant $D$ is to be fixed later.

\subsection{Asymptotic matching.}
Equations (\ref{Z11}) and (\ref{Z21}) give the leading terms in the asymptotic
expansion of the eigenfunction ${\vec \varphi}$ in the regions $x<x_0$ and
$x>-x_0$, respectively.
 Before proceeding to the next order of the expansion
(which will give us the eigenvalue $\lambda$), we need to make sure that
 the two expansions match in the overlap region
 $-x_0<x<x_0$.

In what follows, we will need the asymptotic behaviours of the generalised
eigenvectors ${\dot {\vec \psi}}_N$ and ${\dot {\vec \psi}}_B$.
The solution ${\dot {\vec \psi}}_N$ of
the equation (\ref{Z5})  can be  constructed in quadratures.
The $|\xi| \to \infty$ asymptotic behaviour is straightforward from
the quadrature:
\begin{equation}
{\dot {\vec \psi}}_N (\xi) \rightarrow
- \rho (B+1) e^{-B|\xi|}
\left(
\begin{array}{c} 0 \\ 1
\end{array}
\right),
\label{Z22}
\end{equation}
where the factor
\begin{equation}
\rho=\frac{1}{B(1-B^2)}
\int^{\infty}_{-\infty} {\rm sech\/}^2 \xi (B- {\rm tanh\/} \, \xi) 
e^{B \xi} d \xi 
= \frac{1}{1-B^2} \frac{\pi B/2}{\sin (\pi B/2)}>0.
\label{Z23}
\end{equation}
The $\eta \to \pm \infty$ asymptotics of
the solution ${\dot {\vec \psi}}_B$ of  equation $ {\cal H}_B {\dot {\vec \psi}}_B =-J {\vec \psi}_B'$
are given by
\begin{equation}
{\dot {\vec \psi}}_B(\eta) \rightarrow
\left(
\begin{array}{c}
u_{\pm} e^{-|B \eta-\beta|} \\ w_{\pm} e^{-2|B\eta -\beta|}
\end{array}
\right),
\label{Z24}
\end{equation}
where the constants $u_{\pm}$ and $w_{\pm}$ are easily found by
direct substitution:
\[
u_{\pm} =
\mp \frac{4B(1+B)e^{-\beta}}{4-B^2},  \quad
w_{\pm}= \frac{8B}{4-B^2}.
\label{Z25}
\]

Using (\ref{Z22}),
the $\xi \to +\infty$-asymptotic behaviour 
of the expansion (\ref{Z11}) is given by
\begin{multline}
{\vec \varphi} \rightarrow \left(
\begin{array}{c}
-4a e^{-2(x+s+\beta)} \\  0 \end{array}
\right) + a \lambda \left(
\begin{array}{c} 0 \\
   \rho (B+1) e^{-B(x+s+\beta)}
\end{array}
\right)  \\
+ 8 a \varepsilon e^{B(x+s)}
 \left(
\begin{array}{c} 0 \\
 e^{-2(x+s+\beta)}  \end{array}
\right)
+C_1
\left(
\begin{array}{c} 0 \\
2B(1-B) e^{B(x+s)}
\end{array}
\right),
\label{Z26}
\end{multline}
where we have substituted $x+s+\beta$ for each $\xi$.
The $\eta \to -\infty$-asymptotic behaviour of the expansion (\ref{Z21}) is
\begin{equation}
{\vec \varphi} \rightarrow  b \left(
\begin{array}{c}
4 B e^{2B(x-s) -2 \beta} \\
2 (1+B) e^{-\beta} Be^{B(x-s)-\beta}
\end{array}
\right)
-\lambda b
\left(
\begin{array}{c}
u_-e^{B(x-s)-\beta} \\
v_- e^{2B(x-s)-2\beta}
\end{array}
\right)
+D
\left(
\begin{array}{c}
-4 e^{-2(x-s)-2 \beta}
\\
8  e^{(B-2)(x-s)-2\beta}
\end{array}
\right).
\label{Z27}
\end{equation}
Here we have replaced each $\eta$ with $x-s$.

Equating
 coefficients of the exponential
$e^{Bx}$ in the bottom rows
of (\ref{Z26}) and (\ref{Z27}) determines the constant $C_1$:
\begin{equation}
C_1= b \varepsilon.
\label{ZC}
\end{equation}
Equating
 coefficients of the exponential $e^{-2x}$ in the top rows of
(\ref{Z26})  and (\ref{Z27}) and
 the exponential
$e^{(B-2) x}$ in the bottom rows,
fixes the constant $D$:
 \begin{equation}
 D= a \mu.
 \label{ZD}
 \end{equation}
(Both top and bottom rows lead to equivalent
equations, so only one parameter is fixed.)
Since $D$ turns out to be smaller than $\lambda$ and even $\lambda^2$,
we can drop the term $D {\vec z_1}$ from ${\vec \varphi}_1$
and ${\vec \varphi}_2$.

There is a term in (\ref{Z26}) which does not
have a matching partner in (\ref{Z27}), and the other way
round, there are terms  in (\ref{Z27}) which do not
have counterparts in (\ref{Z26}). Consider, first,
the term proportional to $e^{-B x}$ in Eq.(\ref{Z26}).
 This exponential does not have a partner
in Eq.(\ref{Z27}); however we can add a matching term
$-a \lambda {\dot {\vec \psi}}_N(\xi)$ with
$\xi=\eta+2s+\beta$
to the expansion (\ref{Z21}). This term  will be of order $e^{-4Bs}$
for $\eta \sim 0$  and hence will arise only
at the next, $\varepsilon^2-$,  order of the expansion.
In a similar way, the exponentials $e^{B x}$ in the top row
and $e^{2B x}$ in the bottom row of (\ref{Z27}) do not have counterparts
in the expansion (\ref{Z26}). This can be taken care of by adding the
term $-b \lambda {\dot {\vec \psi}}_B(\eta)$
with $\eta=\xi-2s-\beta$ to Eq.(\ref{Z11}). The top
and bottom rows of this term will appear only
at the  order $\varepsilon^2$ and $\varepsilon^3$, respectively.
Finally, the last unmatched exponential $e^{2 B x}$
in the top row of (\ref{Z27}) will acquire a matching
partner if we add
$\varepsilon^2 {\vec y}_2$ to Eq.(\ref{Z11}), where ${\vec y}_2(\xi)$
is given by Eq.(\ref{Zy2}).

\subsection{Secular equation.}
To identify the constants $a$, $b$ and the eigenvalue $\lambda$ we proceed to
the next order of the perturbation expansion.
As before, we treat the regions $x<0$ and $x>0$ separately.
 For $-\infty< x< x_0$, the order $\varepsilon^2$ gives
\begin{equation}
{\cal H}_N  {\vec \varphi}_2 + {\cal H}_N^{(1)} {\vec \varphi}_1 + a {\cal H}_N^{(2)} {\vec \psi}_N'= \lambda_0 J {\vec \varphi}_1,
\label{Z31}
\end{equation}
where  ${\vec \varphi}_1$ is as in Eq.(\ref{Z11}).
 The solvability condition of Eq.(\ref{Z31})
is
\begin{equation}
\left( {\vec \psi}_N',  {\cal H}_N^{(1)} {\vec \varphi}_1 \right) +
a \left({\vec \psi}_N',  {\cal H}_N^{(2)} {\vec \psi}_N'\right)=
\lambda_0 \left({\vec \psi}_N', J {\vec \varphi}_1 \right),
\label{Z32}
\end{equation}
where $(,)$ stands for the $L^2$-scalar product of two vector-functions
of $\xi$:
\[
\left({\vec f}, {\vec g} \right) = \int_{-\infty}^\infty  \left(
f^1 g^1+ f^2 g^2 \right) d \xi.
\]
To obtain (\ref{Z32}), we have made use of the identity
$({\vec \psi_N}', {\cal H}_N {\vec \varphi}_2)=
({\vec \varphi}_2, {\cal H}_N {\vec \psi_N}')$,
and the fact that ${\cal H}_N {\vec \psi_N}'=0$. However, since ${\vec \varphi}_2$
should include
 terms which have
exponential growth as $\xi \to +\infty$,
the validity of this identity (and hence of the equation (\ref{Z32}))
 may be under suspicion. To reassure that
the solvability condition is indeed correct, we note that the
second-order perturbation should have the
form
\begin{equation}
\varepsilon^2 {\vec \varphi}_2(\xi) =-b \lambda  {\dot {\vec \psi}}_B
 + \varepsilon^2 {\vec y}_2 +C_2 {\vec y}_1 + \varepsilon^2 {\tilde {\vec \varphi}}_2,
\label{Zsol}
\end{equation}
 where $C_2$ is a constant of order $\varepsilon^2$,
 and the function ${\tilde {\vec \varphi}}_2$ is bounded
as $|\xi| \to \infty$. Note that the growing terms in
the right-hand side of (\ref{Zsol}) (the first, second and third terms)
grow no faster than  $e^{2B \xi}$ whereas ${\vec \psi}_N'$ decays as $e^{-2 \xi}$
when $\xi \to +\infty$. Therefore, when doing each of the integrals
 $({\vec \psi}_N', {\cal H}_N {\dot {\vec \psi}}_B)$,
 $({\vec \psi}_N',{\cal H}_N {\vec y}_2)$ and
$({\vec \psi}_N',{\cal H}_N {\vec y_1})$ by parts, the boundary terms vanish.
Consequently, we have
$({\vec \psi}_N', {\cal H}_N {\vec \varphi}_2)=
({\vec \varphi}_2, {\cal H}_N {\vec \psi}_N')$; Q.E.D.

Eq.(\ref{Z32})  can be simplified considerably if we
use the identities arising at the order $\varepsilon^2$ of equations ${\cal H}{\vec \psi}'=0$
and  ${\cal H}{\vec \psi}_s=0$ with ${\cal H}$ expanded as in
(\ref{Z2}) and ${\vec \psi}'$, ${\vec \psi}_s$ as in (\ref{Z6}):
\begin{eqnarray*}
 {\cal H}_N^{(2)} {\vec \psi}_N'+  {\cal H}_N^{(1)}
{\vec \chi}_1
 + {\cal H}_N  {\vec \chi}_2=0,
 \\
 {\cal H}_N^{(1)}  {\vec y}_1
+ {\cal H}_N {\vec y}_2=0.
\end{eqnarray*}
Taking the scalar product with ${\vec \psi}_N'$, these become
\begin{eqnarray}
\left({\vec \psi}_N', {\cal H}_N^{(2)}  {\vec \psi}_N' \right)
+\left({\vec \psi}_N', {\cal H}_N^{(1)} {\vec \chi}_1 \right)=0,
\nonumber
\\
\left(  {\vec \psi}_N', {\cal H}_N^{(1)} {\vec y}_1 \right)=0.
\label{Z35}
\end{eqnarray}
Using Eqs.(\ref{Z7}), (\ref{ZC}) and  (\ref{Z35}), Eq.(\ref{Z32}) simplifies
to
\begin{equation}
a \lambda^2 \left({\vec \psi}_N', J {\dot {\vec \psi}}_N\right) -
b \varepsilon  \lambda   \left( {\vec \psi}_N', J {\vec y}_1 \right)=0.
\label{Z50}
\end{equation}
 Noting that
$\left( {\vec \psi}_N', J {\dot {\vec \psi}}_N \right)= (1/2) {\dot P}_N$
and evaluating the integral $\left({\vec \psi}_N', J {\vec y}_1 \right)$,
  Eq.(\ref{Z50}) becomes, finally,
\begin{equation}
 \frac{\lambda^2}{2} {\dot P}_N a+ 2 \varepsilon \lambda \rho
 B^2 (1-B^2)  \,  e^{-B \beta}  b =0.
\label{Z51}
\end{equation}
Here $P_N$ is the momentum of the N\'eel wall, Eq.(\ref{yN}).

Turning to the region $-x_0< x<\infty$, the eigenfunction has the expansion
\begin{equation}
{\vec \varphi}(\eta)= b {\vec \psi}_B'- \lambda b {\dot {\vec \psi}}_B
+ \varepsilon^2 {\vec \varphi}_2+ {\cal O}(\lambda^3, \mu),
\label{Z80}
\end{equation}
where ${\vec \varphi}_2$ consists of a bounded part ${\tilde {\vec \varphi}}_2$
and a part that grows as $\eta \to -\infty$:
\begin{equation}
\varepsilon^2 {\vec \varphi}_2(\eta)= \lambda^2 {\tilde {\vec \varphi}}_2
- \lambda a {\dot {\vec \psi}}_N(\xi).
\label{Z81}
\end{equation}
The order $\lambda^2$ of the expansion of ${\cal H} {\vec \varphi}=\lambda J {\vec \varphi}$
gives
\begin{equation}
{\cal H}_B \left( \lambda^2 {\tilde {\vec \varphi}}_2 - a \lambda {\dot {\vec \psi}}_N
\right)= -\lambda^2 b J {\dot {\vec \psi}}_B.
 \label{Z71}
 \end{equation}
 The
solvability condition is
\begin{equation}
 - a \lambda \left({\vec \psi}_B', {\cal H}_B {\dot {\vec \psi}}_N
\right)= -\lambda^2 b \left( {\vec \psi}_B', J {\dot {\vec \psi}}_B \right),
 \label{Z72}
 \end{equation}
 where, this time,
 \[
 \left({\vec f},{\vec g}\right) \equiv \int_{-\infty}^{\infty}
 \left( f^1 g^1 +f^2 g^2 \right) d \eta.
 \]

 At first glance, the scalar product on the left-hand side of (\ref{Z72})
 is zero as ${\cal H}_B {\vec \psi}_B'=0$.
 However, ${\dot {\vec \psi}}_N$ grows as $\eta \to -\infty$:
 ${\dot {\vec \psi}}_N \sim e^{-B \eta}$, while ${\vec \psi}_B'$
 does not decay fast enough --- in fact, it decays at exactly the same rate:
 ${\vec \psi}_B' \sim e^{B \eta}$.
 Therefore, $\left({\vec \psi}_B', {\cal H}_B {\dot {\vec \psi}}_N \right)
 \neq \left({\dot {\vec \psi}}_N, {\cal H}_B {\vec \psi}_B' \right)$.
(Note the difference from our analysis of the neighbourhood of
 the N\'eel wall where ${\vec \psi}_N'$ decayed faster than the
 exponentially growing terms in ${\vec \varphi}_2$
 would grow, and hence $\left( {\vec \psi}_N', {\cal H}_N {\vec \varphi}_2 \right)
 = \left( {\vec \varphi}_2, {\cal H}_N  {\vec \psi}_N'\right)$ was indeed true.)
 Making use of the asymptotic expression (\ref{Z22}) and taking care
 of the boundary term, we get, instead:
 \begin{equation}
\left({\vec \psi}_B', {\cal H}_B {\dot {\vec \psi}}_N \right)
= 2  \varepsilon  \rho B^2 (1-B^2) e^{-B \beta}.
 \label{Z73}
 \end{equation}
 Finally, Eq.(\ref{Z72}) becomes
 \begin{equation}
 -2  \varepsilon  \lambda \rho  B^2 (1-B^2) e^{-B \beta}  \,   a
 + \frac{\lambda^2}{2}  {\dot P}_B b=0,
 \label{Z74}
 \end{equation}
where $P_B$ is the momentum of the Bloch wall, Eq.(\ref{yB}).

 Eqs.(\ref{Z51}) and (\ref{Z74}) constitute a system of two linear equations
 for two unknowns, $a$ and $b$. The associated characteristic equation is
 fourth-order in $\lambda$:
 \[
 \lambda^2 \left|
 \begin{array}{lr}
 \lambda {\dot P}_N &
 4 \varepsilon \rho B^2 (1-B^2) \,  e^{-B\beta}   \\
 4 \varepsilon \rho B^2 (1-B^2) \,  e^{-B\beta}   &
 -\lambda {\dot P}_B
 \end{array}
 \right|   \\ =0.
 \label{Z75}
 \]
 The equation has two zero roots corresponding to the translation symmetry
 of the bound state as a whole and variations of the interwall separation;
 there is also a pair of real roots of opposite sign:
\begin{equation}
 \lambda=\pm  2 \pi
 \frac{B^3 e^{-B \beta} }{\sin (\pi B /2)}
 \frac{1}{\sqrt{-{\dot P}_N {\dot P}_B}} e^{-2Bs}.
 \label{Z76}
 \end{equation}
(Note that ${\dot P}_N$ and ${\dot P}_B$ are opposite in sign --- see Fig.\ref{momentum_vs_V};
hence we have a positive quantity under the square root).
This is the first  result of our analysis. The second result
is the relation between $a$ and $b$, the coefficients of the
leading terms in the expansion of ${\vec \varphi}$:
\begin{equation}
\frac{a}{b} =  \pm \sqrt{
-\frac{{\dot P}_B}{{\dot P}_N}}.
 \label{Z77}
 \end{equation}
Here the plus corresponds to the positive eigenvalue
and minus to the negative one. (That is, the ``unstable" eigenfunction
has $a$ and $b$ of the same sign, both positive or negative, whereas the
``stable" eigenfunction has the coefficients of the opposite sign.)
   In what follows it will be convenient to choose the normalisation
   such that $a=\sqrt{{\dot P}_B}>0$ for either sign of $\lambda$, while
 $b=\sqrt{-{\dot P}_N}>0$ for $\lambda>0$ and
$b=-\sqrt{-{\dot P}_N}<0$ for $\lambda <0$. Thus, the
normalised eigenfunctions are:
\begin{subequations} \label{noconfusion}
\begin{eqnarray}
{\vec \varphi}_u = \sqrt{{\dot P}_B}
\left( {\vec \psi}_N' -|\lambda| {\dot {\vec \psi}}_N +
\varepsilon {\vec \chi_1} \right)
+  \sqrt{-{\dot P}_N} \varepsilon {\vec y}_1 +{\cal O}(\varepsilon^2),
\nonumber  \\
{\vec \varphi}_s = \sqrt{{\dot P}_B}
\left( {\vec \psi}_N' + |\lambda| {\dot {\vec \psi}}_N +
\varepsilon {\vec \chi_1} \right)
-  \sqrt{-{\dot P}_N} \varepsilon {\vec y}_1 +{\cal O}(\varepsilon^2)
\nonumber \\
\label{noconfusion_a}
\end{eqnarray}
for $x<0$, and
\begin{eqnarray}
{\vec \varphi}_u = \sqrt{-{\dot P}_N}
\left( {\vec \psi}_B' -|\lambda| {\dot {\vec \psi}}_B \right)
 +{\cal O}(\varepsilon^2), \nonumber \\
{\vec \varphi}_s = -\sqrt{-{\dot P}_N}
\left( {\vec \psi}_B' + |\lambda| {\dot {\vec \psi}}_B \right)
 +{\cal O}(\varepsilon^2)
 \label{noconfusion_b}
\end{eqnarray}
\end{subequations}
for $x>0$.

We conclude our calculation by commenting on the case of the
complex involving a {\it left}-handed Bloch wall
($\sigma=+1$) and  on the situation where the Bloch wall
(right- or left-handed) is on the
left of the N\'eel wall (i.e. $s<0$).
These cases do not require a separate treatment. Indeed, let
$(u(x), w(x))$ be the eigenvector associated with the eigenvalue
$\lambda$ (positive or negative) of the bubble (\ref{Family}) with $s>0$ and
$\sigma=-1$. Changing $\sigma \to -\sigma$ changes the sign
of the off-diagonal entries of the matrix (\ref{LinearisedOperator});
hence the vector $(u(x), -w(x))$ will give  the eigenfunction associated
with the eigenvalue $-\lambda$ for the bubble with $\sigma=1$.
Therefore, the ``unstable" eigenvector for the bubble
with $s>0$ and $\sigma=+1$, obtains from the ``stable" eigenvector
of the bubble with $s>0$, $\sigma=-1$, just by changing the sign
of its bottom component.
To find the eigenfunctions for the Bloch wall on the left of the
N\'eel, we note a symmetry of the bubble solution (\ref{Family}):
$\psi_b(x,-s)=\psi_b^*(-x,s)$.
According to this symmetry, changing $s \to -s$ in Eq.(\ref{Family}) is
equivalent to replacing $x \to -x$ and taking
 complex conjugate. Consequently,
 the vector $(u(-x),-w(-x))$ will serve as
the eigenfunction associated with the eigenvalue $-\lambda$
for the bubble with $\sigma=-1$  and $s<0$. Finally, $(u(-x),w(-x))$ and $\lambda$
are the eigenvector and the associated eigenvalue pertaining to
the situation with $\sigma=+1$, $s<0$.

\subsection{Numerical verification}

\begin{figure}
\includegraphics[ height = 2.1in, width = 0.5\linewidth]{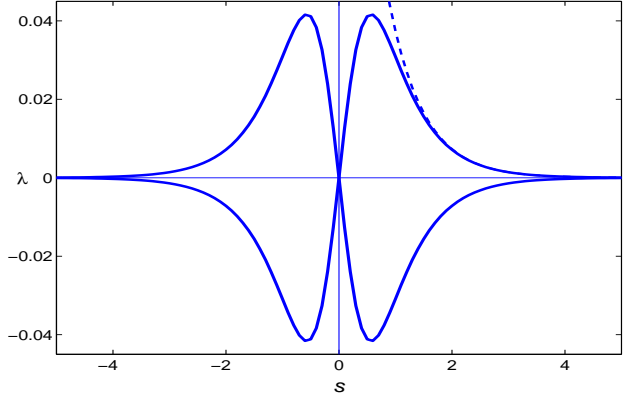}
\caption{\sf Two discrete real eigenvalues of the eigenvalue
problem (\ref{EVprob}). The solid lines depict eigenvalues found
numerically while the dashed line gives the perturbation
approximation (\ref{Z76}). This plot is for the case where the
Bloch wall bound in the complex is right-handed;  changing the
chirality of the Bloch wall switches the two branches, but leaves
the overall shape the same. This figure pertains to $h = 0.2$; for
other $h$, the functions $\lambda(s)$ look similar. }
\label{Spectrum}
\end{figure}

The above
perturbation results were verified numerically.
 (We used Fourier expansions of $\vec{\varphi}$\  over $120$\
positive and $120$\ negative harmonics, on the interval $[-40, \, 40]$.)
In agreement with our expectations, for $s=\infty$ we have found four
zero eigenvalues. Two of these remain at zero as $s$ is decreased
from infinity.
The other two zero eigenvalues move to the real axis; one
becomes positive and the other negative (Fig.\ref{Spectrum}).
 Their magnitudes are equal. This is a general property of linearizations of Hamiltonian systems of
the form (\ref{EVprob}), with $\mathcal{H}$\ hermitian and $J$\ as in (\ref{vectorthings});
it follows from the conservation of symplectic
areas \cite{Arnold}.
The eigenvalues return to zero as $s \to 0$.
These are the only
eigenvalues with nonzero real part
in the spectrum of $\mathcal{H}$; hence for any finite, nonzero $s$, there is exactly one unstable mode.
In addition, there are two pure imaginary eigenvalues detaching from the continuous
spectrum as $s$\ decreases from infinity. These do not reach the origin; they remain on the imaginary axis
and hence do not cause instability.

In addition to the numerical solutions of the eigenvalue problem
(\ref{EVprob}), we performed numerical simulations of the
full time-dependent nonlinear Schr\"odinger equation
(\ref{NLS2}).
Simulations
were carried out
 using a split-step pseudospectral method under periodic 
 boundary conditions (see e.g. \cite{Herbst}).
Typically, we used an interval $(-L/2, \, L/2)$\ with $L = 60$,
although for small $h$\
and large $s$\ (i.e. when the walls bound
in the bubble are far apart and decay slowly to the background), $L = 120$\ was necessary. The  timestep was set at
$\Delta t = 2.5\times 10^{-3}$. The code is stable for
$\Delta t < \pi^{-1} (L/N)^2$, so
for $L =120$ we were able to use $N = 1024$ modes, while for $L = 60$, we were limited to
$N = 512$. The spatial resolution $\Delta x =2\pi L/N$ was the same in both cases.
Results of numerical simulations are presented in the next two sections,
concurrently with predictions of the asymptotic
analysis.

\begin{figure}
\includegraphics[ height = 2.1in, width = 0.5\linewidth]{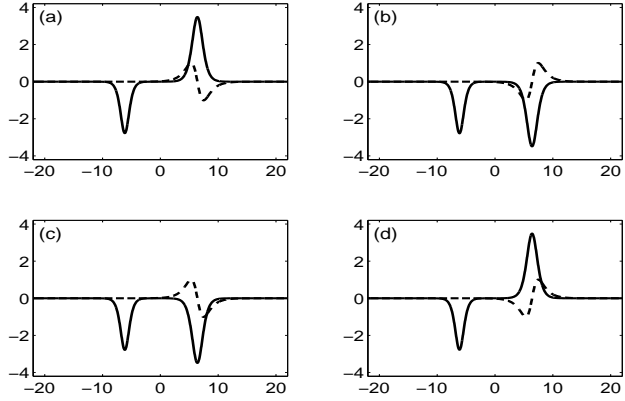}
\caption{\sf  The  eigenfunctions of the bubble involving the
right-handed [$\sigma=-1$, (a,b)]  and left-handed [$\sigma=+1$, (c,d)]
Bloch wall. The eigenfunction in (a,c) is associated with the
positive  and that  in (b,d) with the
negative eigenvalue.
 In this figure, $s = 5$\ and  $h = 0.2$. The real
part is shown by the solid line, while the imaginary part is dashed.
}
\label{Evectors2}
\end{figure}

\section{Interpretation of the eigenfunction}
\label{Interpretation}

The eigenfunction
associated with the  eigenvalue $\lambda>0$
admits interpretation in terms
of the motions of the constituent walls. Let, for definiteness,
$s>0$ and $\sigma=-1$, and note that the
dissociating bound state can be described by
\begin{subequations}
\label{03}
\begin{equation}
{\vec \psi}(x,t) =
{\vec \psi}_N \left( v_N(t); x+s+ \delta s_N(t) -x^{(0)} \right) +...
\end{equation}
in the region $x<0$, and
\begin{equation}
{\vec \psi}(x,t) =
{\vec \psi}_B  \left( v_B(t); x-s- \delta s_B(t) -x^{(0)} \right) +...
\end{equation}
\end{subequations}
for $x>0$.
Here ${\vec \psi}_N (v;x)$  and ${\vec \psi}_B (v;x)$
are stationary solutions of Eq.(\ref{stationary}) with small $v$
obtained by continuation from ${\vec \psi}_N (x)$  and ${\vec \psi}_B (x)$,
respectively, and ``..."  includes the part of the
perturbation which cannot be reduced to the translation and velocity
boost of the corresponding wall. Note that since we have not specified
this part, the parameters $v_N$, $v_B$, $\delta s_N$, and $\delta  s_B$
are not defined uniquely in (\ref{03}); to fix  these parameters
we need to restrict the ``..."-part in some way.
To do this, we note that Eq.(\ref{03}) can be represented as
${\vec \psi}_b(x)+ \delta {\vec \psi}(x,t)$, where ${\vec \psi}_b(x)=({\cal R}, {\cal I})$
is the stationary bubble (\ref{Family}) with some $s$ and $x^{(0)}$, and a small perturbation
$\delta {\vec \psi}(x,t)$ can be  written in the form
\begin{subequations}
 \label{04}
 \begin{eqnarray}
\delta {\vec \psi}(x,t) =
v_N(t) {\dot {\vec \psi}}_N
 + \delta s_N(t) {\vec \psi}_N'
+{\vec \nu}(x,t),
  \\
 \delta {\vec \psi}(x,t) =  v_B(t) {\dot {\vec \psi}}_B
 -  \delta s_B(t) {\vec \psi}_B' +{\vec \nu}(x,t)
 \end{eqnarray}
\end{subequations}
in $x<0$ and $x>0$, respectively.
Here
\begin{eqnarray*}
{\dot {\vec \psi}}_N  \equiv
\left. \partial_v {\vec \psi}_N (v; x+s-x^{(0)}) \right|_{v=0}; \\
{\vec \psi}_N'  \equiv
\partial_x {\vec \psi}_N (0; x+s-x^{(0)}); \\
          {\dot {\vec \psi}}_B  \equiv
\left. \partial_v {\vec \psi}_B (v; x-s-x^{(0)})\right|_{v=0}; \\
{\vec \psi}_B'  \equiv
\partial_x {\vec \psi}_B (0; x-s-x^{(0)}).
\end{eqnarray*}
(We remind that the overdot stands for the partial derivative w.r.t. $v$, not $t$, here.)
For the given $s$ and $x^{(0)}$, we can fix  $v_N$, $v_B$, $\delta s_N$, and $\delta  s_B$
by requiring that the ``remainder'' ${\vec \nu}(x,t)$ be $J$-orthogonal to the subspace
spanned by the velocity boosts and translations:
\begin{subequations}
\label{ortho_4}
\begin{eqnarray}
\int_{-\infty}^0  \langle {\vec \nu} ,  J {\dot {\vec \psi}}_N \rangle
\, dx =
\int_{-\infty}^0  \langle {\vec \nu} ,  J {\vec \psi}_N' \rangle
\, dx =0, \label{ortho_4a} \\
\int_0^{\infty}  \langle {\vec \nu} ,  J {\dot {\vec \psi}}_B \rangle
\, dx=
\int_0^{\infty}  \langle {\vec \nu} ,  J {\vec \psi}_B' \rangle
\, dx =0. \label{ortho_4b}
\end{eqnarray}
\end{subequations}
Here $\langle, \rangle$  denotes the scalar product in the two-dimensional
Euclidean space: $\langle {\vec f},  {\vec g} \rangle \equiv
f_1 g_1+f_2 g_2$.
From (\ref{ortho_4}) we have
\[
v_N= \frac{(\delta {\vec \psi}, J {\vec \psi}_N')}
         {({\dot  {\vec \psi}}_N, J {\vec \psi}_N')}
\]
and similar expressions for $v_B$, $\delta s_N$, and $\delta  s_B$.
One more consequence of the constraints (\ref{ortho_4}) is that
${\vec \nu}(x,t)$ is linearly independent from ${\dot {\vec \psi}}_N$ and
${\vec \psi}_N' $ in the region $x<0$ and from  ${\dot {\vec \psi}}_B$
and ${\vec \psi}_B'$ in the region $x>0$. (Indeed, if we assumed
${\vec \nu}=c_1 {\vec \psi}_N' +c_2 {\dot {\vec \psi}}_N$ and
substituted this in (\ref{ortho_4a}), then using the fact that
$({\vec \psi}_N', J {\dot {\vec \psi}}_N)=\frac12 {\dot P_N} \neq 0$,
we would immediately get $c_1=c_2=0$.)

The perturbation (\ref{04}) can be expanded over 
solutions of the equation
(\ref{EVprob}):
\begin{multline}
\delta {\vec \psi}= {\cal M}_u e^{\lambda t} {\vec \varphi}_u + {\cal M}_s e^{-\lambda t} {\vec \varphi}_s \\
+ {\cal N}_0 {\vec \psi}_b'  + {\tilde {\cal N}_0}  \partial_s {\vec \psi}_b
+ {\cal Q}^{(+)}  e^{i \omega_0 t} {\vec \phi}^{(+)} + {\cal Q}^{(-)} e^{-i \omega_0 t}  {\vec \phi}^{(-)}  \\
+ \int_{-\infty}^{\infty} {\cal Q}^{(1)}(k) e^{i \omega t} {\vec \phi}^{(1)}_k dk +
\int_{-\infty}^{\infty} {\cal Q}^{(2)} (k) e^{i \omega t}   {\vec \phi}^{(2)}_k dk.
\label{expansion1}
\end{multline}
In Eq.(\ref{expansion1}), ${\vec \varphi}_u$ and ${\vec \varphi}_s$
are the eigenfunctions
associated with discrete real eigenvalues $\lambda$ and $-\lambda$,
respectively, where we set $\lambda>0$;
${\vec \psi}_b'$ and  $\partial_s {\vec \psi}_b$ are the two zero modes
while
 ${\vec \phi}^{(+)}$ and ${\vec \phi}^{(-)}$
are eigenfunctions associated with pure imaginary discrete eigenvalues $\pm i \omega_0$.
Finally, ${\vec \phi}_k^{(1)}$ and $ {\vec \phi}_k^{(2)}$ are solutions of
the continuous spectrum,
${\cal H} {\vec \phi}_k^{(1,2)}=i \omega J  {\vec \phi}_k^{(1,2)}$, with
 $\omega^2= \frac14 (k^2+4)(k^2+B^2)$,
and $\omega(k) >0$ for $k>0$ and $ \omega(k) <0$ for $k<0$.

As $t$ grows, the expansion (\ref{expansion1}) tends to ${\cal M}_u e^{\lambda t} {\vec \varphi}_u$,
where ${\vec \varphi}_u (x)$ is given by (\ref{noconfusion}).
This should be identified with the
 large-$t$ behaviour of (\ref{04}).
Using linear independence of  vectors
${\vec \psi}_N'$,  ${\dot {\vec \psi}}_N $,
${\vec \nu}$ in $x<0$ and vectors  ${\vec \psi}_B'$,  ${\dot {\vec \psi}}_B$,
${\vec \nu}$ in $x> 0$,
we get
\begin{eqnarray}
v_N(t) \to - \lambda {\cal M}_u \sqrt{{\dot P}_B}  e^{\lambda t}, \nonumber \\
v_B(t) \to -\lambda {\cal M}_u \sqrt{-{\dot P}_N}   e^{\lambda t};
\label{vNB} \\
\delta s_N(t) \to {\cal M}_u \sqrt{{\dot P}_B}    e^{\lambda t}, \nonumber \\
 \delta s_B(t) \to - {\cal M}_u \sqrt{-{\dot P}_N}
 e^{\lambda t}
\label{sNB}
\end{eqnarray}
as $t \to \infty$.
Equations (\ref{vNB}) and (\ref{sNB}) are consistent in the
sense that the velocity of the wall determined from the deformation
of its shape coincides with the velocity defined by the
position of its centre:
\[
v_N(t) \to  -\frac{d}{dt} \delta  s_N(t), \quad
v_B(t) \to \frac{d}{dt} \delta  s_B(t).
\]

\begin{figure*}
\includegraphics[height = 2.0in, width = 0.45\linewidth]{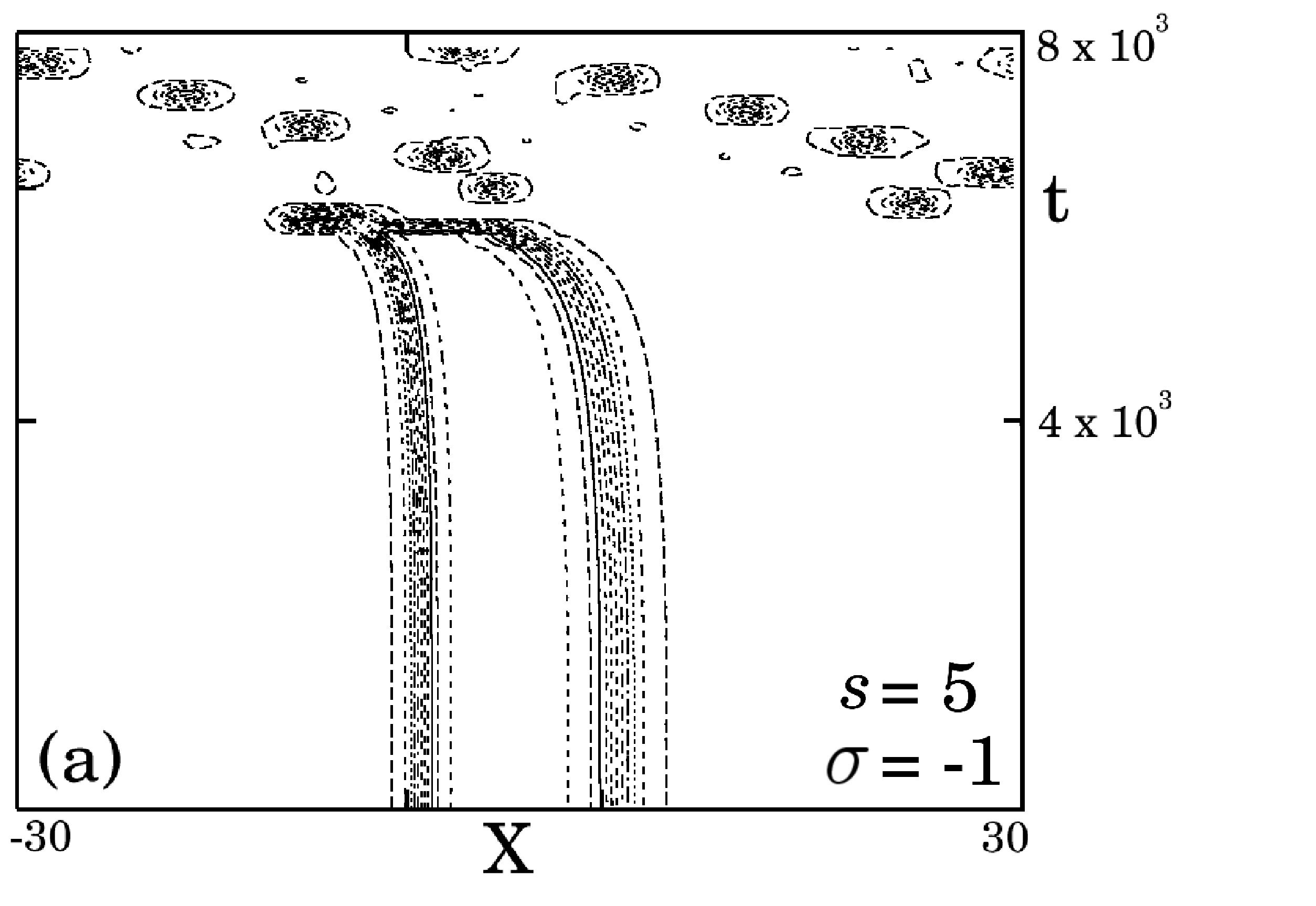}
\includegraphics[height = 2.0in, width = 0.45\linewidth]{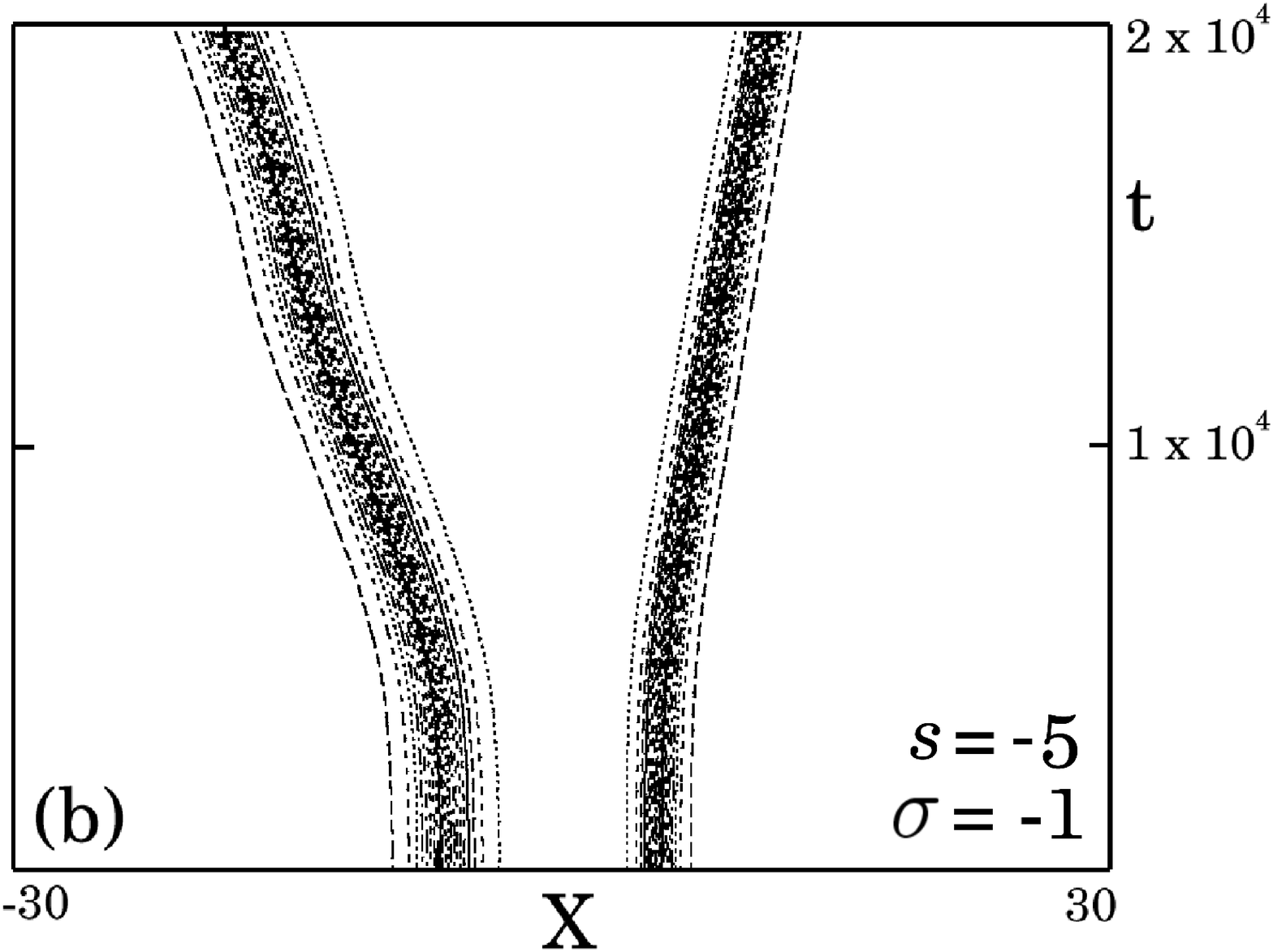}
\includegraphics[height = 2.0in, width = 0.45\linewidth]{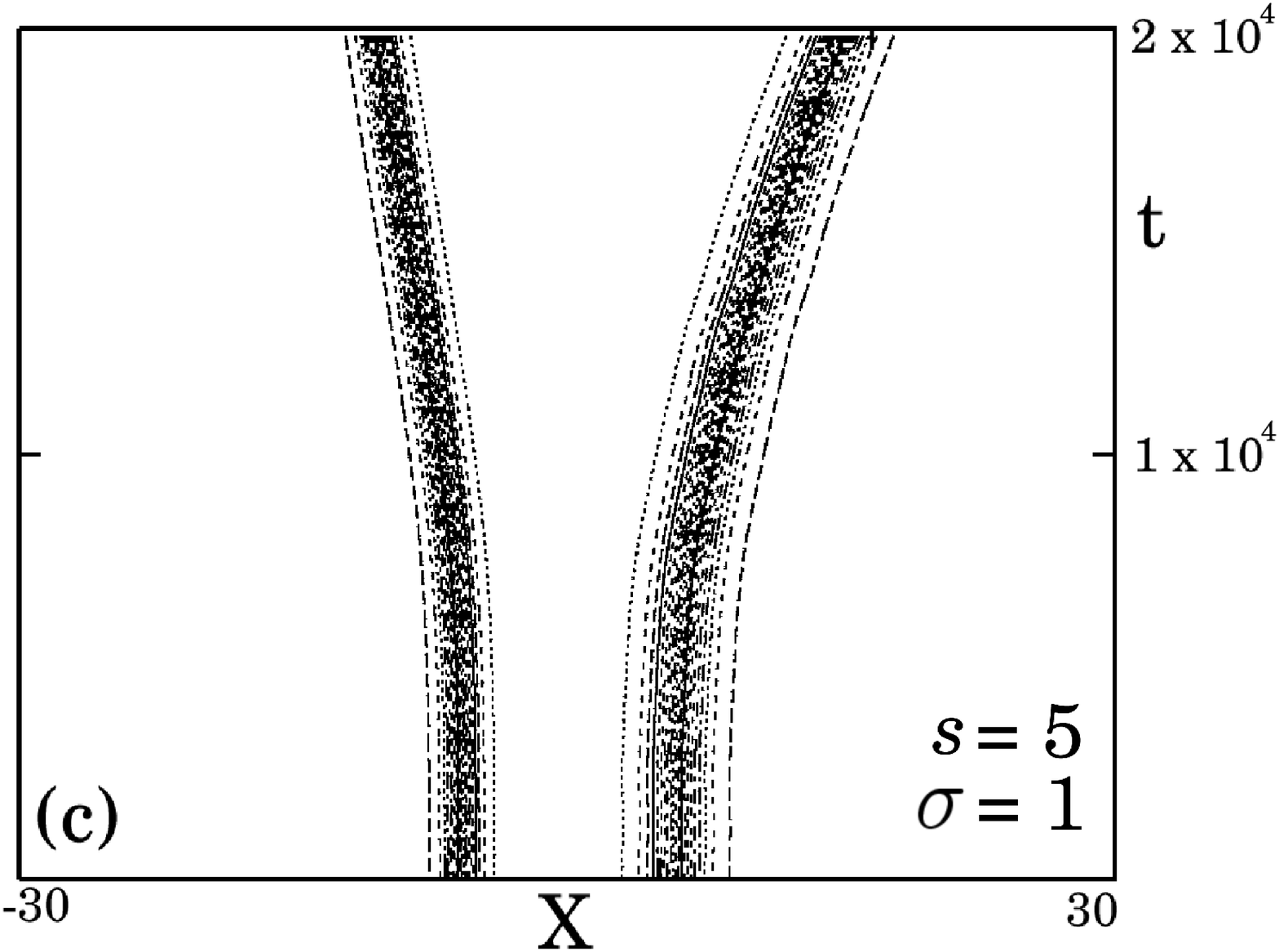}
\includegraphics[height = 2.0in, width = 0.45\linewidth]{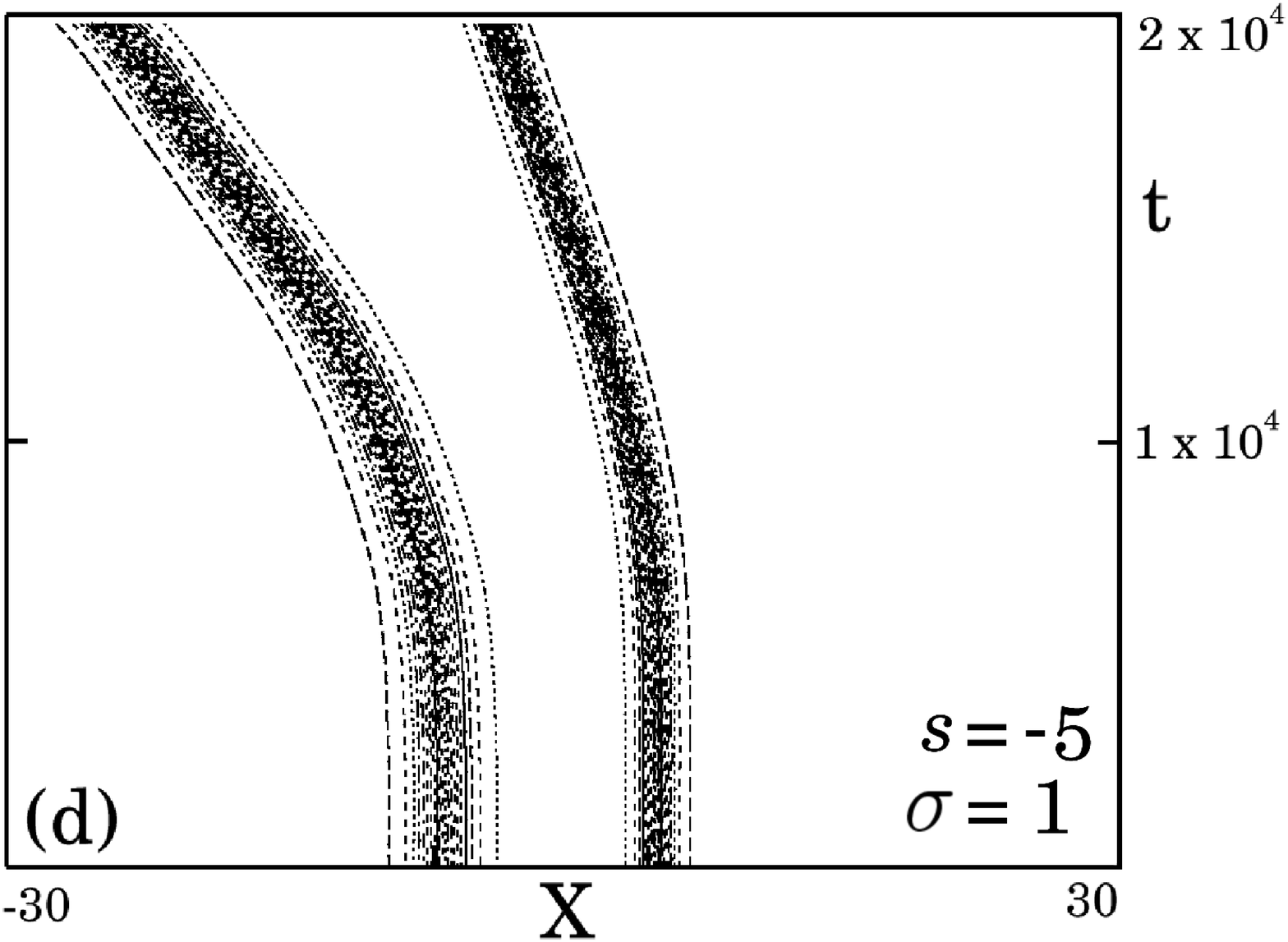}
\caption{\sf The development of the instability of
the bubble (\ref{Family}).
Shown are the curves of constant real part.
The wide and narrow trails pertain to the Bloch and N\'eel walls, respectively.
Panels (a) and (b) correspond to bound states involving
right-handed Bloch walls ($\sigma=-1$), with $s =
5$\ and $s = -5$, respectively. The evolution of the bubbles with the
left-handed Bloch walls ($\sigma=+1$) is depicted in panels
(c) and (d), also with $s =
5$\ and $s = -5$, respectively.
No perturbations were added to the stationary
initial condition (\ref{Family}) ``by hand"; the initial
(uncontrollable) disturbance was merely due to the
discretisation of the equation. The collision of two
walls in (a) produces a breather.
This breather  moves so quickly that the contour plot seems to indicate the
presence of two separate breathers
whereas there is in fact only one.
This effect is due to a sparse time-sampling
of the profile which we had to resort to for better
visualisation.
The image in (a)
has been generated by saving a profile once only every $200$\ time units.
In these plots, $h = 0.1$.
}
\label{BubPics1}
\end{figure*}

Note that the derivative ${\dot P}_N$ is greater, in absolute
value, than ${\dot P}_B$ (see Fig.\ref{momentum_vs_V}).
Consequently, Eq.(\ref{vNB})
implies that the Bloch wall arising from the dissociation of the bubble,
always moves faster than the emerging N\'eel wall:
$|v_B|> |v_N|$.
The same  Eq.(\ref{vNB}) implies that the
corresponding velocities and accelerations
of  the Bloch and N\'eel walls will be colinear.
Consequently, the walls
emerging from the decay of the bubble
with $s>0$ and $\sigma=-1$ will be
moving in the same direction.
(The direction will of course be determined by the initial perturbation.)
This conclusion is in agreement with direct numerical
simulations of Eq.(\ref{NLS2}). (See Fig. \ref{BubPics1}(a)).

\begin{figure}
\includegraphics[height = 2.0in, width = 0.5\linewidth]{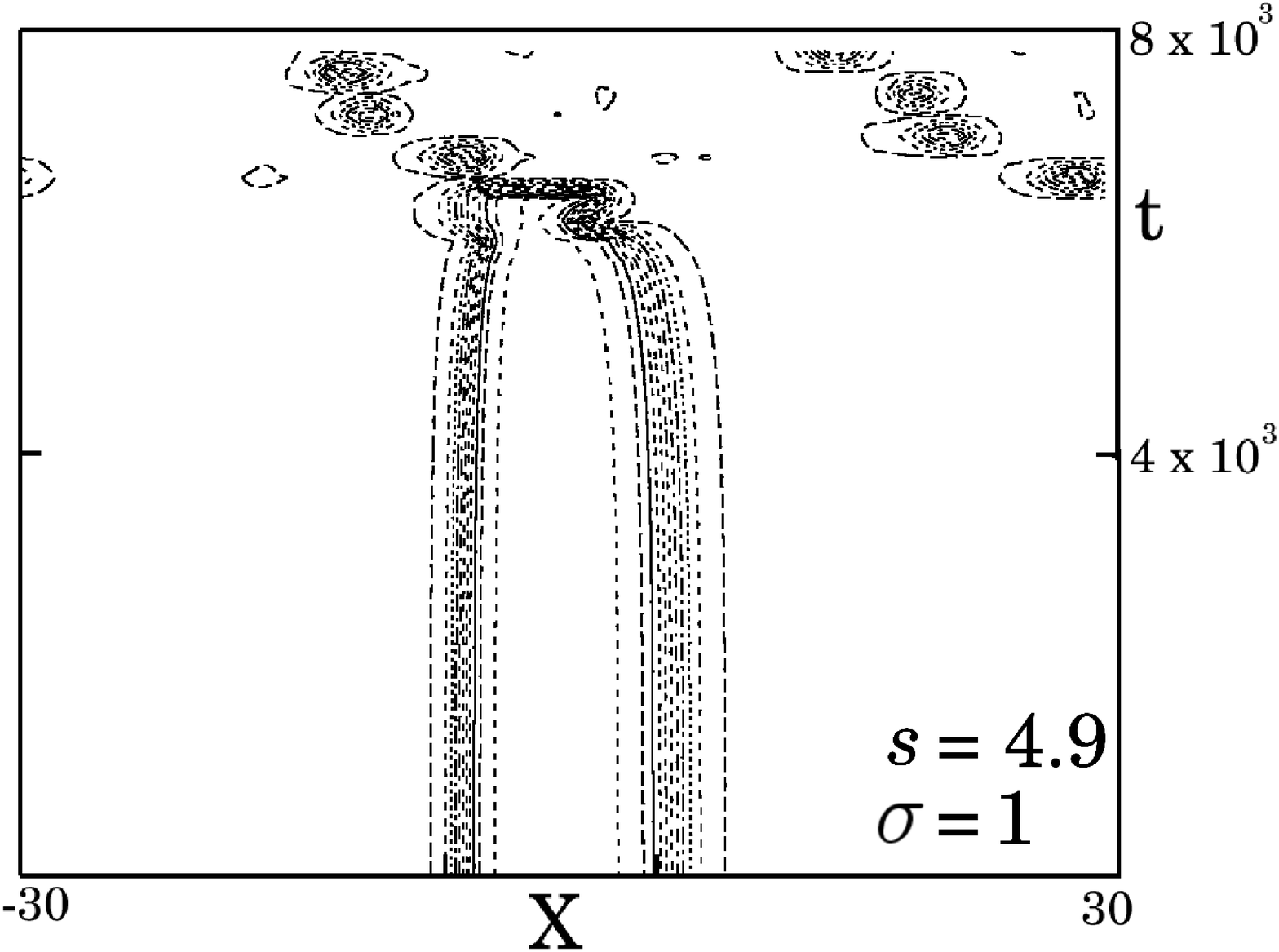}
\caption{\sf  Evolution of the bound state involving a  left-handed Bloch wall
($\sigma=+1$), with  $s = 4.9$.
This initial condition is almost identical to the initial condition that led to Fig.\ref{BubPics1}(c); however,
the direction
of motion of the walls
 is now exactly opposite.
The reason, of course, is that ${\cal M}_u$ was
positive in the perturbation of
Fig.\ref{BubPics1}(c), whereas here, the ``unstable" eigenfunction
${\vec \varphi}_u$  is excited
with a negative coefficient.
As in Fig.\ref{BubPics1}, in this plot $h = 0.1$.
}
\label{BubPics2}
\end{figure}

It follows then from equation
(\ref{vNB})  that if ${\cal M}_u >0$, the two
walls will be moving to the left, with the Bloch wall
catching up with the N\'eel wall.
This was indeed seen in simulations (Fig. \ref{BubPics1}(a)).
(In our numerical simulations, the choice ${\cal M}_u >0$ was accidental; the
initial perturbation of the bubble was entirely due to the discretisation
errors and hence beyond our control.)
 On the contrary,  if ${\cal M}_u <0$,
the walls will be moving to the right, with the
N\'eel wall lagging behind.

Changing $\sigma \to - \sigma$ swaps around the top
components of the eigenfunction ${\vec \varphi_u}$
associated with $\lambda>0$ and eigenfunction
 ${\vec \varphi_s}$ pertaining to the negative eigenvalue $-\lambda$.
Therefore,
the ``unstable" eigenfunction of the bubble
with $\sigma=+1$ involves coefficients $a$ and $b$
of the {\it opposite\/} signs.
As a result, the fragments of its decay --- the
N\'eel and left-handed Bloch wall --- will be moving in opposite
directions.
The simulations confirm this (Fig.\ref{BubPics1}(c)).

As we have already mentioned, the choice of the coefficient ${\cal M}_u$
was beyond our control in the simulations. For $s =5$
and $\sigma=+1$, our
discretisation induced ${\cal M}_u >0$. However, for a slightly different value of $s$,
{\it viz.\/} $s=4.9$, the ``unstable" eigenfunction ${\vec \varphi_u}$
was seen to be excited
with the coefficient ${\cal M}_u <0$ (Fig.\ref{BubPics2}).

 The effect of the change $s \to -s$ amounts to
changing $\sigma \to -\sigma$ and $x \to -x$; therefore,
the dissociating bound state of the
right-handed Bloch
wall on the left of   the N\'eel wall
 will  produce walls
moving in opposite directions. [Simulation shown in
Fig.\ref{BubPics1}(b).]
Finally, a left-handed Bloch wall ($\sigma=+1$) and a N\'eel wall placed on its right
(i.e. $s<0$) will move in the same direction. If (as it happened in our simulations)
${\cal M}_u$ is $>0$, the Bloch will be leaving the N\'eel wall behind
[Fig.\ref{BubPics1}(d)].

In the case of converging walls, the result of the
collision in all cases is the formation of a spatially localised,
temporally oscillating object (which we refer to as a breather),
propagating over the constant background
$\psi =1$.  An asymptotic expression for the breather was derived in the
previous submission \cite{BWZ1};
here, we simply note that it persisted
indefinitely after it was formed in our simulations.

\section{The Bloch-N\'eel Interaction}
\label{Interaction}

\subsection{The method.}

Our knowledge of the phase space in the vicinity of the unstable bound state
of the Bloch and N\'eel walls can be used to make conclusions on the
evolution of  particular initial configurations of the two walls. Any initial condition comprising
well separated Bloch and N\'eel walls can be represented as
\[
\psi(x,0)= \psi_b(x) + \delta \psi(x,0),
\]
where $\psi_b(x)$ is the bubble (\ref{Family}) with some suitably chosen
separation $s$, centred at some point $x^{(0)}$, and $\delta \psi(x,0)$
is a small perturbation.
This initial perturbation can be expanded as in (\ref{expansion1})
where we just need to set $t=0$.
In order to find the coefficient ${\cal M}_u$ with which the unstable eigenvector
${\vec \varphi}_u $ enters the perturbation, we note that the $J$-product
of the vector ${\vec \varphi}_s $ with any solution
of Eq.(\ref{EVprob}) except ${\vec \varphi}_u $
is zero:
\begin{eqnarray*}
({\vec \varphi}_s, J {\vec \varphi}_s)=
({\vec \varphi}_s, J {\vec \psi_b}')
=({\vec \varphi}_s, J \partial_s {\vec \psi_b} ) \\ =
({\vec \varphi}_s, J {\vec \phi}^{(\pm)} )
=({\vec \varphi}_s, J {\vec \phi}^{(1)}_k )=
({\vec \varphi}_s, J {\vec \phi}^{(2)}_k )=0.
\end{eqnarray*}
[This is a simple consequence of the hermiticity of the
operator ${\cal H}$ in Eq.(\ref{EVprob}):
if ${\cal H} {\vec \zeta}_1=\lambda_1 J {\vec \zeta}_1$ and ${\cal H} {\vec \zeta}_2=\lambda_2 J {\vec \zeta}_2$,
then $(\lambda_1+\lambda_2) ({\vec \zeta}_1, J{\vec \zeta}_2)=0$. Hence
$({\vec \zeta}_1, J{\vec \zeta}_2)=0$ unless $\lambda_1=-\lambda_2$.]
Therefore,
\begin{equation}
{\cal M}_u= \frac{({\vec \varphi}_s, J \delta {\vec \psi} )}
{({\vec \varphi}_s, J {\vec \varphi}_u )}.
\label{Mplus}
\end{equation}
The sign of ${\cal M}_u$ determines the direction of colinear motion of the two walls.
In fact, the denominator in (\ref{Mplus}) is
positive and the sign of ${\cal M}_u$ is determined just by the
sign of $({\vec \varphi}_s, J \delta {\vec \psi})$.
Indeed, using the representation (\ref{noconfusion_a}) for ${\vec \varphi}_s(x)$,
  ${\vec \varphi}_u(x)$
and invoking the identity (\ref{Z50}), one can check that the
integral over the negative semiaxis is zero to the leading order
in $\varepsilon$:
\[
\int_{-\infty}^0  \langle {\vec \varphi}_s ,  J {\vec \varphi}_u \rangle
\, dx =\mathcal{O}(\varepsilon^2).
\]
We remind that $\langle, \rangle$  is the scalar product in the two-dimensional
Euclidean space: $\langle {\vec f},  {\vec g} \rangle \equiv
f_1 g_1+f_2 g_2$
whereas $(,)$ stands for the $L^2$-scalar product: $\left({\vec f}(x),
{\vec g}(x)\right) =
\int_{-\infty}^{\infty} \langle {\vec f}, {\vec g} \rangle \, dx$.
On the other hand, using the expansion (\ref{noconfusion_b}) for $0<x<\infty$,
we obtain
\[
\int_0^{\infty}  \langle {\vec \varphi}_s,  J {\vec \varphi}_u \rangle
\, dx = -\lambda \, {\dot P}_N \, {\dot P}_B,
\]
which is positive for $\lambda>0$ (see Fig.\ref{momentum_vs_V}).
Thus, $({\vec \varphi}_s, J {\vec \varphi}_u )=
-\lambda  \, {\dot P}_N \, {\dot P}_B
+ \mathcal{O}(\varepsilon^2) >0$. Q.E.D.

\subsection{Example.}

As a characteristic example, we consider the initial condition of the
form
\begin{equation}
\psi(x,0)= - \psi_N(x+x_1) \psi_B(x-x_1)
\label{producto}
\end{equation}
with some $x_1>0$.
In applications, this product function is often used as an approximation
for two well-separated dark solitons, in the same
way as two
weakly overlapping bright solitons are usually approximated
by their sum. (See e.g. our previous submission \cite{BWZ1} where
 this type of ansatz was employed for the variational analysis of the
N\'eel-N\'eel and Bloch-Bloch interactions.)
Assuming that the Bloch wall is right-handed and that $x_1$ is large, we have
\begin{subequations}
\label{product}
\begin{eqnarray}
\mbox{Re} \, \psi(x,0)= -\mbox{tanh} \, \xi
+ 2  e^{2B(\xi-2 x_1)} \, \mbox{tanh} \,  \xi   +...,  \nonumber \\
\mbox{Im} \,  \psi(x,0)= 2 \mbox{sech} \beta \,
e^{B(\xi-2 x_1)}   \, \mbox{tanh} \, \xi +...
\end{eqnarray}
in the region $x<0$, and
\begin{eqnarray}
\mbox{Re} \, \psi(x,0)= \mbox{tanh} (B \eta-\beta)
 -2 \, e^{-2(\eta +2x_1-\beta/B)} \, \mbox{tanh} (B \eta-\beta)
+...,  \nonumber \\
\mbox{Im} \, \psi(x,0)= \mbox{sech} \beta  \, \mbox{sech}
(B \eta-\beta)
-2 \, \mbox{sech} \beta \, e^{-2(\eta +2x_1-\beta/B)} \,
\mbox{sech} (B \eta-\beta) +...
\end{eqnarray}
\end{subequations}
in the region $x>0$.
Here we have defined $\xi$ and $\eta$ according to
\begin{equation}
\xi= x+x_1, \quad \eta  =x-x_1 + \frac{\beta}{B}.
\label{xi_eta_1}
\end{equation}
Note that the leading, $\mathcal{O}(\varepsilon^0)$,
terms in (\ref{product}) coincide, formally, with the leading terms
in the asymptotic expansions of the bubble, Eqs.(\ref{Z1}) and (\ref{Z120}).
However the definitions of $\xi$ and $\eta$ in (\ref{xi_eta_1})
will not, in general, be consistent with the definitions of $\xi$ and $\eta$
for the bubble, i.e. Eqs.(\ref{1star}) and (\ref{2stars}).
To achieve the consistency,  we introduce an
additional translation parameter
$x^{(0)}$ in the definitions  (\ref{1star}) and (\ref{2stars}):
\begin{equation}
\xi= x-x^{(0)}+s+\beta, \quad \eta =x-x^{(0)}-s.
\label{xi_eta_2}
\end{equation}
Thus we allow for translations of the bubble to ensure
the coincidence of the leading terms in (\ref{product}) and (\ref{Z1}), (\ref{Z120}).
Equating (\ref{xi_eta_1}) to (\ref{xi_eta_2}) we obtain 
parameters of such a reference bubble:
\begin{equation}
s=x_1-\frac{\beta}{2} \left( 1+ \frac{1}{B} \right),
\quad
x^{(0)}= \frac{\beta}{2} \left( 1- \frac{1}{B} \right).
\label{010}
\end{equation}

Subtracting the bubble solution with parameters (\ref{010})
from the initial condition (\ref{producto}), we obtain the
following expressions for
$\mbox{Re} \, \delta \psi \equiv \mbox{Re} \, \psi(x,0) - \mathcal{R}(x)$
and $\mbox{Im} \, \delta \psi \equiv \mbox{Im} \, \psi(x,0)- \mathcal{I}(x)$:
\begin{subequations}
\begin{eqnarray}
\mbox{Re} \, \delta \psi(x) =\varepsilon^2 e^{-2B \beta}
\sinh(2 \beta)
\frac{e^{2(B-1) \xi}}{\cosh^2 \xi}+...,
\nonumber  \\
\mbox{Im} \,  \delta \psi(x)=
2 \varepsilon B e^{-B \beta} \frac{e^{(B-1) \xi}}{\cosh \xi}+...
\label{014}
\end{eqnarray}
in the region $x<0$, and
\begin{multline}
\mbox{Re} \,  \delta \psi(x)
=\mu e^{-2 \beta}
\sinh(2 \beta)
\frac{e^{2(B-1) \eta}}{\cosh^2 (B \eta-\beta)}+...,
 \\
\mbox{Im}  \, \delta \psi(x)=  
2 \mu  e^{-2 \eta} \, \frac{\cosh (B \eta) -
e^{-2 \beta} \mbox{sech} \beta \,\cosh (B \eta-\beta) }{\cosh^2 (B \eta-\beta)}
+... \\
\end{multline}
\end{subequations}
for $x>0$.
Here $\varepsilon=e^{-2Bs}$ and $\mu=e^{-4s}$; ${\cal R}$ and ${\cal I}$ are the
real and imaginary parts of the solution (\ref{Family}): $\psi_b= {\cal R}+i{\cal I}$.

In order to evaluate the integral
$({\vec \varphi}_s, J \delta {\vec \psi})$ in Eq.(\ref{Mplus}),
we note that
the positive semiaxis of $x$ gives a contribution of the
order $e^{-(2+B)s}$. This is exponentially smaller than
the contribution of the negative semiaxis (which is of the
order $e^{-2Bs}$) and hence can be neglected. Using
(\ref{noconfusion_a}) and (\ref{014}), we get then
\[
({\vec \varphi}_s, J \delta {\vec \psi}) =
2 \varepsilon
\sqrt{-{\dot P}_N}
 B e^{-B \beta} \int_{-\infty}^{\infty}
e^{(B-1) \xi} \mbox{sech}^3 \xi d \xi+ ...
\]
which is positive. Hence ${\cal M}_u>0$, and,
according to Eqs.(\ref{vNB})-(\ref{sNB}), both walls will
be moving to the left.

\begin{figure*}
\includegraphics[height = 2.0in, width = 0.45\linewidth]{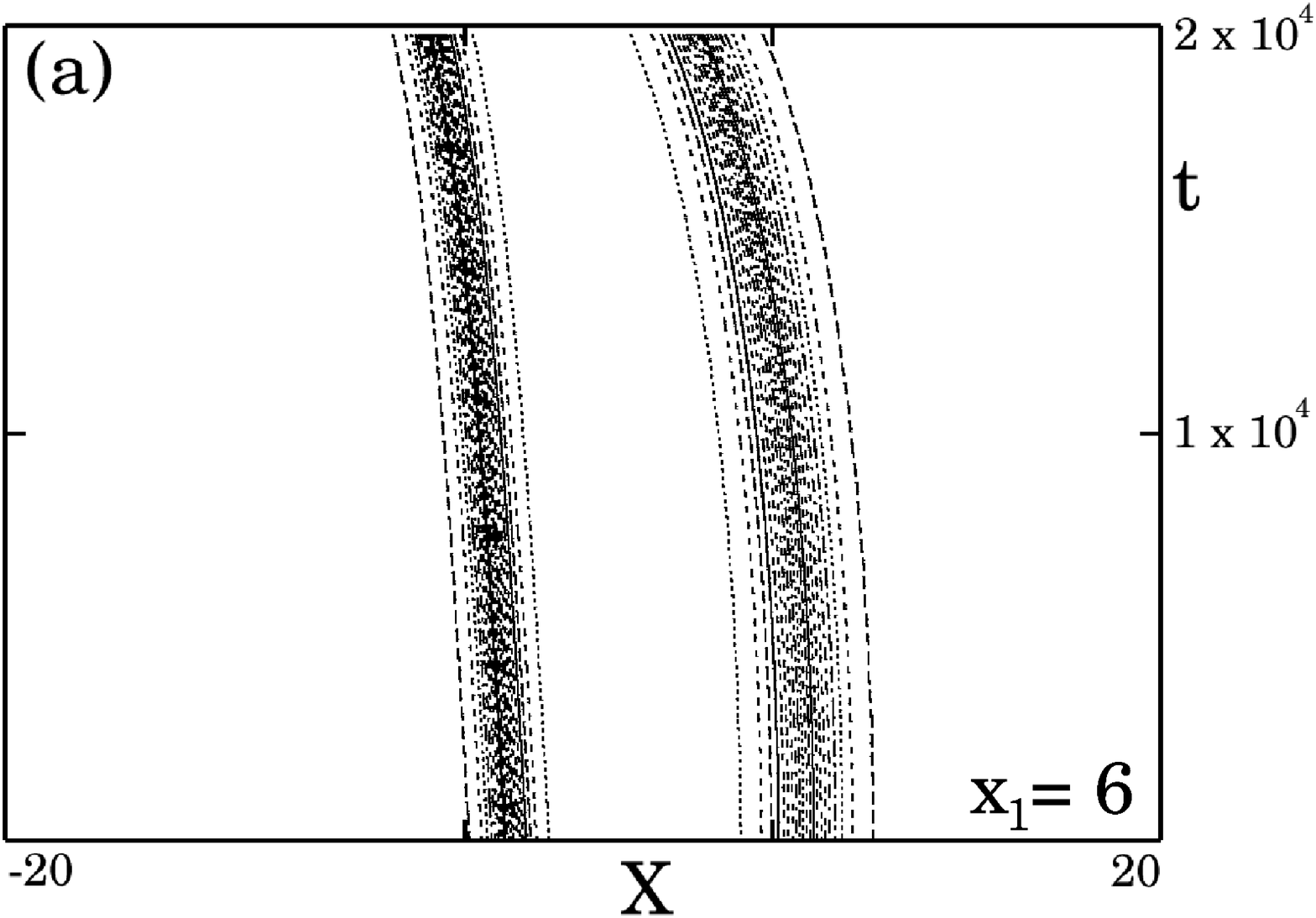}
\includegraphics[height = 2.0in, width = 0.45\linewidth]{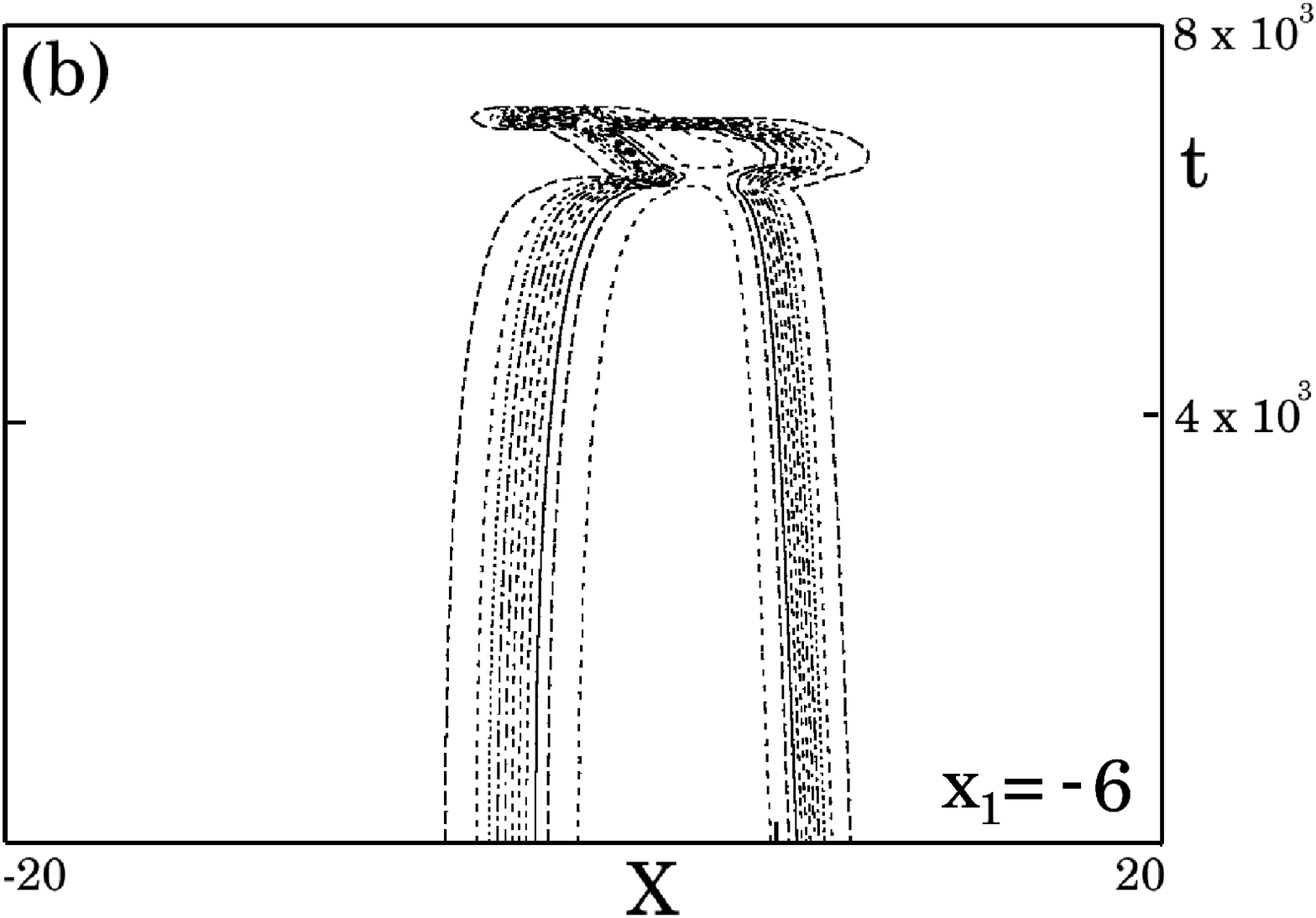}
\includegraphics[height = 2.0in, width = 0.45\linewidth]{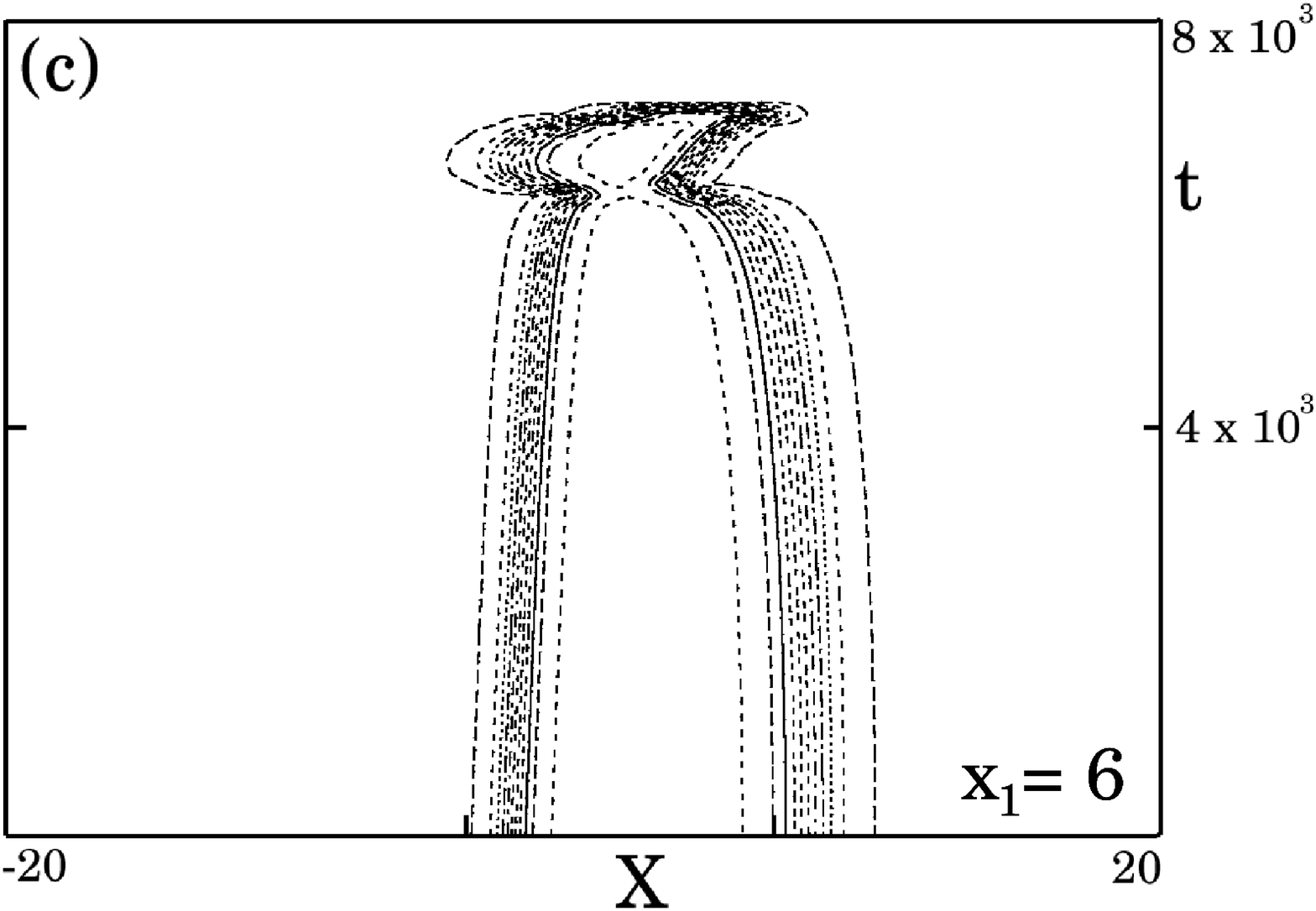}
\includegraphics[height = 2.0in, width = 0.45\linewidth]{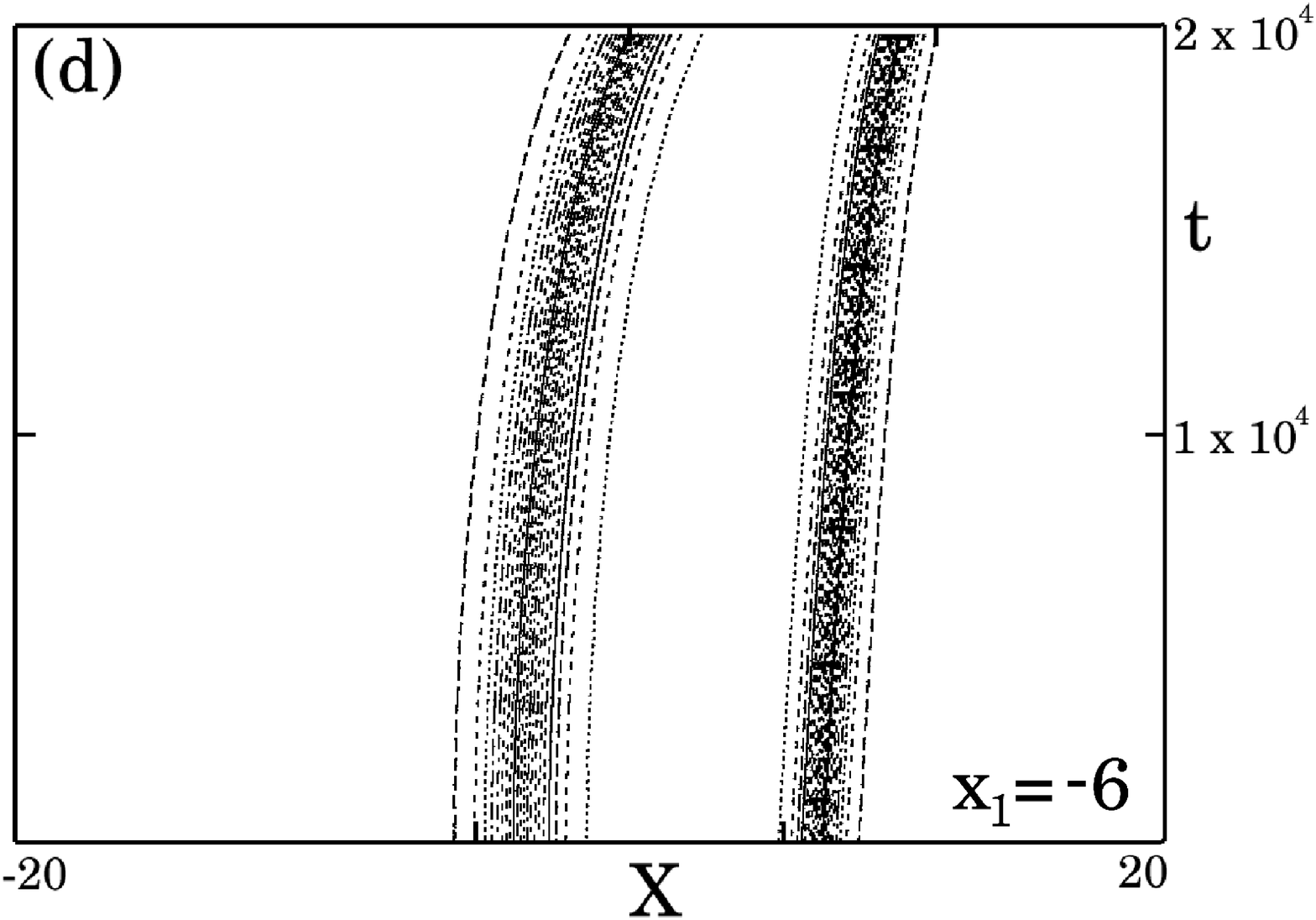}

\caption{\sf  Typical results of the evolution of an initial condition
of the form (\ref{producto}). Panels (a) and (b) correspond to
the
{\it right}-handed Bloch wall, to the right ($x_1 > 0$)  and the left ($x_1 < 0$)
of the N\'eel wall, respectively. Panels (c) and (d) correspond to the {\it left}-handed
Bloch wall, to the right and the left of the N\'eel wall,
respectively. The Bloch wall moves towards the N\'eel wall in all cases. The N\'eel wall moves towards
or away from the Bloch wall
depending on whether it is on the left or right of the Bloch wall, and on
the Bloch wall's chirality. After the collisions in (b) and (c), a fast-moving breather is
formed; however, for visual clarity, this solution is not shown. Here, $h = 0.1$.
}
\label{BubPics3}
\end{figure*}

The direct numerical simulations of Eq.(\ref{NLS2})
verify these conclusions. Fig.\ref{BubPics3}
presents  simulations of the initial condition (\ref{producto})
with $x_1= \pm 6$ and $h=0.1$, for both chiralities of the
Bloch wall.
 (These results are representative  of all $h$\ provided $|x_1|$\ is
sufficiently large).
 When the Bloch wall is right-handed and $x_1$ is positive,
i.e. when the Bloch wall is on the right of the N\'eel, both walls
move to the left (Fig.\ref{BubPics3}(a)) ---  precisely as our analysis predicted.
Using symmetries of the initial condition (\ref{producto}) and the
evolution equation (\ref{NLS2}), one can readily check that
the situations shown in Fig.\ref{BubPics3}(b-d) are in
agreement with the asymptotic analysis as well.

\section{Concluding Remarks}
\label{Conclusions}

 \subsection{Energy considerations}

Equation (\ref{NLS2}) is conservative, with the energy integral given by
\begin{equation}
E = \frac 12 \int \left\{  |\psi_x|^2 +  |\psi|^4 - \frac{2}{A^2} |\psi|^2 
\right. 
\left. - \frac{h}{A^2} \left[\psi^2 + (\psi^*)^2\right] + 1 \right\} \, dx. 
\label{Energy}
\end{equation}
The bubbles, being  time-independent
solutions of Eq.(\ref{NLS2}),  must render the functional
(\ref{Energy}) stationary:
$\delta E = 0$. Since they depend on $s$\
smoothly, it follows that $dE/ds = 0$:
the energy of the bound state is independent of the distance between its constituents.
This  seems to suggest  that the binding
energy of the   Bloch and N\'eel walls is zero, implying the non-interaction of the walls.

However, our analysis has revealed that the walls do, in fact, interact. We have
shown that all bound states
with the parameter $s \neq 0$\ are  exponentially
unstable against the decay into constituent walls. Exponential growth of the intersoliton separation
(see Eq.(\ref{sNB})) is a clear
manifestation of a nonvanishing
interaction between the walls --- for noninteracting walls, the separation would only grow linearly.
As for the zero binding energy, it can be reconciled with the nonvanishing interaction
by noticing that the bound N\'eel wall acquires a small static
 imaginary part (see Eq.(\ref{Z1})).  (The bound Bloch wall also acquires
a stationary  perturbation,  but it is exponentially smaller
than the perturbation of the N\'eel wall.)
To leading order, the force between two walls is made up of
the force between their real parts and the force between their
imaginary parts while the force between the real part of one wall and imaginary
part of the other one, is of the second order of smallness \cite{MalNep}.
The imaginary part of any nontrivial stationary solution  decays
as $\mathcal{O}(e^{-Bx})$, while
the real part decays
as  $\mathcal{O}(e^{-2Bx})$\ or $\mathcal{O}(e^{-2x})$.
Consequently, the force caused by the overlap of the imaginary parts of the
soliton tails is  stronger than the force caused by overlapping real parts. Thus
although the amplitude of the
imaginary excitation of the N\'eel wall is exponentially small for large $s$, the interaction of
this excitation with (the imaginary part of) the Bloch wall is enough to balance
the force between the Bloch wall and the ``naked" N\'eel wall.
If we decrease $s$, and the two walls in the complex are pulled closer,
the increase of (the absolute value of) the
binding energy of their real parts is offset by storing more
energy of the opposite sign in the imaginary excitations; the total energy remains invariant.
A similar mechanism was described by Ostrovskaya {\it et al} \space  in the context of  the dark-bright  solitons
of the undriven vector nonlinear Schr\"odinger equation \cite{gluons}; there, the
imaginary excitations were referred to as ``solitonic gluons''.

\subsection{Walls as particles}

Next, we need to explain, qualitatively, the anomalous behaviour
apparent in Figs.\ref{BubPics1} (a,d), and \ref{BubPics3} (a,d),
where the velocities and accelerations of the two interacting walls are seen to have the
same (rather than the opposite) directions. The key observation here is
that the Bloch and N\'eel walls, considered as point-like particles,
have masses of opposite signs (see Fig.\ref{momentum_vs_V}). One can
easily conceive a simple model system of positive- and negative-mass particles,
which exhibits the observed phenomenology. Let $x_1$ and $x_2$ be the
coordinates of the two particles, with $z \equiv x_1-x_2$, and let $m_1>0$ and
$m_2<0$ be their masses. Consider the Hamiltonian
\begin{equation}
H= \frac{1}{2m_1} (p_1-f_1)^2+ \frac{1}{2m_2} (p_2-f_2)^2,
\label{0H}
\end{equation}
where $p_1,p_2$ are the momenta of the particles, and the functions $f_1=f_1(z)$ and
$f_2=f_2(z)$ decay, exponentially, as $|z| \to \infty$. The equations of
motion are
\begin{subequations} \label{NP}
\begin{eqnarray}
{\dot p}_1= -\frac{\partial H}{\partial x_1}=
\frac{1}{m_1} (p_1-f_1) f_1' + \frac{1}{m_2} (p_2-f_2) f_2', \label{NP1} \\
{\dot p}_2= -\frac{\partial H}{\partial x_2}= -{\dot p}_1, \label{NP2} \\
{\dot x}_1= \frac{\partial H}{\partial p_1}=
\frac{1}{m_1} (p_1-f_1), \label{NP3} \\
{\dot x}_2= \frac{\partial H}{\partial p_2}
= \frac{1}{m_2} (p_2-f_2), \label{NP4}
\end{eqnarray}
\end{subequations}
where the overdot  indicates differentiation with respect
to time ($t$) (and not $v$ as earlier in the text).
 The prime stands for the derivative w.r.t. $z$.

For any $z=z_0$, the equations (\ref{NP}) have a fixed point, describing an unstable
bound state of two particles:
\begin{equation}
p_1=f_1(z_0), \quad p_2=f_2(z_0).
\label{0BS}
\end{equation}
This fixed point is an analogue of the bound state of two walls, Eq.(\ref{Family}).
Linearising Eqs.(\ref{NP}) about the equilibrium  point (\ref{0BS}), and letting
$\delta p_{1,2}(t), \delta x_{1,2}(t) \propto e^{\lambda t}$, we get a pair of nonzero
real eigenvalues
\begin{equation}
\lambda^{(\pm)} = \pm \frac{1}{\sqrt{-m_1 m_2}} [f_1'(z_0)+ f_2'(z_0)].
\label{elam}
\end{equation}
(There is also a pair of {\it zero\/} eigenvalues resulting from the overall
translations $x_{1,2} \to x_{1,2}-x^{(0)}$ and the freedom in choosing $z_0$.)
Working out the associated eigenvectors and substituting them into the
linearised equations (\ref{NP3})-(\ref{NP4}), we find
\begin{subequations} \label{mdx}
\begin{eqnarray}
m_1 \delta {\dot x}_1= \pm \frac{K^{(\pm)}}{\sqrt{|m_2|}} \delta z, \label{mdxa} \\
|m_2| \delta {\dot x}_2=  \frac{K^{(\pm)}}{\sqrt{m_1}} \delta z, \label{mdxb}
\end{eqnarray}
\end{subequations}
where $K^{(\pm)}$ are coefficients dependent on $m_{1,2}$ and $f_{1,2}'(z_0)$.

If $f_1'(z_0)+f_2'(z_0)>0$, the unstable eigenvalue is $\lambda^{(+)}$, and
we  keep the top sign in (\ref{mdxa}). This is the case of anomalous behaviour:
the velocities $\delta {\dot x}_1$ and $\delta {\dot x}_2$
as well as accelerations
$\delta {\ddot x}_1$ and $\delta {\ddot x}_2$
have the same sign and so the particles move in the same direction, like
the walls in Figs.\ref{BubPics1} (a,d), and \ref{BubPics3} (a,d).   If
$f_1'(z_0)+f_2'(z_0)<0$, the positive eigenvalue is $\lambda^{(-)}$ and the
``unstable" eigenvector is given by the
bottom equation in (\ref{mdxa}). In this case the two particles
have  opposite velocities and opposite accelerations; this corresponds
to the ``normal" interaction of the Bloch and N\'eel wall seen in
the panels (b,c) of Figs.\ref{BubPics1} and \ref{BubPics3}.
(It is fitting to note here that the anomalous behaviour does not
mean that the two particles violate the third Newton's law.
The force exerted by the particle 2 on particle 1, $-\partial H/\partial x_1$,
is exactly opposite to the back reaction force, $-\partial H/\partial x_2$.)

\subsection{Conclusions}

Finally, we summarise the results of this investigation. 

The essence of our approach is to consider the interacting 
Bloch and N\'eel walls as a perturbation of their unstable
stationary complex. The interaction between the two walls is
characterised by the eigenfunction associated with the positive eigenvalue
in the spectrum of the linearised operator
(evaluated at this stationary solution).

Using matched
asymptotic expansions, we have evaluated the real eigenvalues for the
Bloch-N\'eel complex
 and constructed the associated eigenfunctions. 
The structure of the ``unstable" eigenfunction for the complex consisting
of a right-handed Bloch wall on the right, and a N\'eel wall on the left,
indicates that the walls emerging from the decay of this ``bubble", 
will be moving
colinearly, i.e.
in the same direction. 
This rule determines the evolution 
of a general configuration of a right-handed Bloch wall on the right and a N\'eel 
wall on the left --- as long as the walls are sufficiently far away 
from each other. 
Using symmetry properties of the Bloch-N\'eel complex, 
we can also predict the type of motion (colinear or antilinear)
for other chiralities and mutual arrangements of the walls. 
The asymptotic analysis is in agreement with direct 
numerical simulations of the interacting Bloch and
N\'eel walls.

Although the unstable eigenfunction determines the type
of motion of two interacting walls,  knowing the 
structure of this eigenfunction is insufficient to 
know which direction this motion will take.
(Depending on whether the unstable eigenfunction is excited with
a positive or negative coefficient, the colinearly moving walls may 
travel to the left or to the right. Similarly, 
in those cases where the unstable eigenfunction sets the opposite
direction of motion for the walls, they may travel either
towards or
away from each other.)
The actual
direction of the colinear or antilinear motion depends on the 
particular initial
condition and can be determined by the projection of the 
corresponding perturbation of the 
stationary complex 
on its unstable eigendirection. We have evaluated this projection for an
initial condition in the form of a product of the Bloch- and N\'eel-wall
solutions,
and verified the conclusions of the asymptotic analysis numerically.

Finally, we have interpreted the anomalous
interaction of the Bloch and N\'eel walls as a dynamics of
two interacting particles, one with positive and the other with negative mass.

\acknowledgments
It is a pleasure to thank Elena Zemlyanaya
for testing, numerically, some aspects of the perturbation theory 
in section \ref{Splitting}.
We are grateful to Vladimir Gerdzhikov
 and Boris Malomed for useful discussions.
One of the authors (IB) thanks Dr Reinhard Richter
and Prof Ingo Rehberg for their hospitality at
the University of Bayreuth where this project was completed.
IB is a Harry Oppenheimer
Fellow;  also supported by the NRF of South Africa under grant 2053723.
SW was supported by the NRF
of South Africa.

\end{document}